\documentclass[11pt,preprint]{aastex}

\newcommand{\bd}{BD+44$^\circ$493}
\newcommand{\Teff}{$T_\mathrm{eff}$}
\newcommand{\logg}{$\log g$}
\newcommand{\kms}{km~s$^{-1}$}
\usepackage{comment}

\shorttitle{The Ninth Magnitude Carbon-Enhanced Metal-Poor Star \object{\bd}}
\shortauthors{Ito et al.}

\begin{document}

\title{Chemical Analysis of the Ninth Magnitude Carbon-Enhanced Metal-Poor Star \object{\bd}\altaffilmark{1}} 

\author{Hiroko Ito\altaffilmark{2,3}, Wako Aoki\altaffilmark{2,3}, Timothy C. Beers\altaffilmark{4,5}, Nozomu Tominaga\altaffilmark{6,7}, Satoshi Honda\altaffilmark{8}, and Daniela Carollo\altaffilmark{9,10}}

\altaffiltext{1}{Based on data collected at the Subaru Telescope, which is operated by the National Astronomical Observatory of Japan.}
\altaffiltext{2}{Department of Astronomical Science, School of Physical Sciences, The Graduate University for Advanced Studies (SOKENDAI), 2-21-1 Osawa, Mitaka, Tokyo 181-8588, Japan} %; hiroko.ito@nao.ac.jp.}
\altaffiltext{3}{National Astronomical Observatory of Japan (NAOJ), 2-21-1 Osawa, Mitaka, Tokyo 181-8588, Japan; aoki.wako@nao.ac.jp.}
\altaffiltext{4}{National Optical Astronomy Observatory, Tucson, AZ 85719: beers@noao.edu}
\altaffiltext{5}{Department of Physics \& Astrophysics, Center for the Study of Cosmic Evolution, and Joint Institute for Nuclear Astrophysics (JINA), Michigan State University, East Lansing, MI 48824-1116}
\altaffiltext{6}{Department of Physics, Faculty of Science and Engineering, Konan University, 8-9-1 Okamoto, Kobe, Hyogo 658-8501, Japan; tominaga@konan-u.ac.jp.}
\altaffiltext{7}{Institute for the Physics and Mathematics of the Universe (IPMU), the University of Tokyo, 5-1-5 Kashiwanoha, Kashiwa, Chiba 277-8569, Japan.}
\altaffiltext{8}{Kwasan Observatory, School of Science, Kyoto University, Kyoto, 377-0702, Japan; honda@kwasan.kyoto-u.ac.jp.}
\altaffiltext{9}{Department of Physics and Astronomy, Astronomy, Astrophysics \&
Astrophotonic Research Center, Macquarie University, North Ryde, NSW
2109, Australia; daniela.carollo@mq.edu.au}
\altaffiltext{10}{INAF-Osservatorio Astronomico di Torino, Strada Osservatorio 20, Pino Torinese, 10020, Torino, Italy}

\begin{abstract}
We present detailed chemical abundances for the bright carbon-enhanced
metal-poor (CEMP) star {\bd}, previously reported on by Ito et al. Our
measurements confirm that {\bd} is an extremely metal-poor
([Fe/H]$=-3.8$) subgiant star with excesses of carbon and oxygen. No
significant excesses are found for nitrogen and neutron-capture
elements (the latter of which place it in the CEMP-no class of stars).
Other elements that we measure exhibit abundance patterns that are
typical for non-CEMP extremely metal-poor stars. No evidence for
variations of radial velocity have been found for this star. These
results strongly suggest that the carbon enhancement in {\bd} is
unlikely to have been produced by a companion asymptotic giant-branch
star and transferred to the presently observed star, nor by pollution
of its natal molecular cloud by rapidly-rotating, massive, mega
metal-poor ([Fe/H] $< -6.0$) stars. A more likely possibility is that
this star formed from gas polluted by the elements produced in a
``faint'' supernova, which underwent mixing and fallback, and only
ejected small amounts of elements of metals beyond the lighter
elements. The Li abundance of {\bd} ($A$(Li)=$\log$(Li/H)+12$=1.0$) is
lower than the Spite plateau value, as found in other metal-poor
subgiants. The upper limit on Be abundance ($A$(Be)
=$\log$(Be/H)+12$<-1.8$) is as low as those found for stars with
similarly extremely-low metallicity, indicating that the progenitors
of carbon- (and oxygen-) enhanced stars are not significant sources of
Be, or that Be is depleted in metal-poor subgiants with effective
temperatures of $\sim$5400~K.

\end{abstract}

\keywords{Galaxy: abundances --- stars: abundances --- stars: individual(\bd) --- stars: Population II}

\section{Introduction}

The measured chemical abundances of extremely metal-poor (EMP; [FeH]
$< -3.0$) stars are believed to reflect the early chemical enrichment
of the Universe, in particular the yields of the first generations of
massive stars. Stars at even lower metallicities (e.g., [Fe/H]$<-3.5$)
exhibit other distinct characteristics. Among these: (1) The
metallicity distribution function (MDF) shows evidence for a drop at
[Fe/H]$\sim -3.5$ \citep[e.g., ][]{Yong12}, though not as sharp a
cutoff as that suggested by some previous studies \citep{schorck09};
(2) The fraction of carbon-enhanced metal-poor (CEMP) stars is quite
high, on the order of 40\% according to Beers \& Christlieb 2005, and
other recent determinations; and (3) there exist a number of
``outliers'' exhibiting chemical abundance ratios that are not often
seen for relatively more metal-rich stars \citep[excess of Mg, deficiency of
Si, etc.; see ][]{Ryan96, Norris01, Johnson2002, Cayrel04, Cohen08,
Lai08, Yong12}.

%Ryan et
%al. 1996; Norris et al. 2001; Johnson 2002; Cayrel et al. 2004; Cohen et
%al. 2008; Lai et al. 2008; Yong et al. 2012).

Although great progress has been made over the past few decades by
searches for very metal-poor stars in the halo \citep[see ][for a 
recent review]{Ivezic12}, the number of stars known to have
metallicities below [Fe/H] = $-3.5$ is still rather small, probably
reflecting the nature of the halo system's MDF.  For example, the most
recent update of the SAGA database for chemical abundances of
metal-poor stars (\citet{Suda08} includes less than 30 objects
with [Fe/H] $ < -3.5$; a number of others have recently been added by
\citet{Yong12} and \citet{Aoki12}). This makes the discovery of
the nature of {\bd} rather remarkable \citep{Ito09}, given that it is
some {\it three magnitudes} brighter than the next example of such
extreme stars. As shown by \citet{Ito09}, this star is a subgiant with
[Fe/H]$ = -3.8$, exhibits clear over-abundances of carbon and oxygen,
and a lack of enrichment among the neutron-capture elements, placing
it within the sub-class of CEMP-no stars (see Beers \& Christlieb
2005).  Thanks to its relative brightness, the abundances of other
important elements that can constrain the possible progenitors of this
star can be determined.  For example, here we report a strong upper
limit for Be, which is measurable only in the near-UV spectral region.
The Be abundance for this star, as well as the abundance of Li, are
useful for constraining mixing processes that might have occured in
this star, and/or the formation process of these light elements in the
early Universe.  The rather low upper limit we obtain for lead ($\log
\epsilon$(Pb)$< -0.10$) is of particular note, as previous predictions
have suggested that, for stars as low in [Fe/H] as {\bd}, one would
have expected log$\epsilon$(Pb) $\sim +1.5$ \citep{Cohen06} from s-process.

This paper is outlined as follows. In \S~\ref{sec:obs} we report details
of the observations, reductions, and measurements of the spectral data,
as well as related data used for our study, including derived
kinematics. The determination of stellar parameters and chemical
abundance analyses are reported in \S~\ref{sec:param} and
\S~\ref{sec:ana}. In \S~\ref{sec:disc}, implications of the
abundance results for {\bd} derived by our analyses are discussed,
focusing on the origin of carbon and oxygen excess and the light elements
Li and Be. 

\section{Observations and Measurements}\label{sec:obs}

\subsection{High-Resolution Spectroscopy}

\object{\bd} was observed with the HDS (High Dispersion Spectrograph;
\citealt{Noguchi02}) at the Subaru Telescope in 2008 August, October,
and November, during an open-use intensive program (PI: W. Aoki) for
high-resolution follow-up spectroscopy of very and extremely metal-poor
stars discovered with the SDSS/SEGUE survey \citep{Yanny09}. In addition
to targets from SEGUE, some brighter very metal-poor candidates were
also observed during twilight. The star \object{\bd} was first observed
with a 10 minute exposure at dawn of 2008 August 22 (Hawaii Standard
Time). The grating setting covered 4030-6730\,\AA, and the slit width
was 0.6 arcsec (a $2\times2$ on-chip binning was applied), yielding a
resolving power of $R \sim 60,000$. Our initial abundance analysis of
this spectrum surprisingly indicated that the metallicity of the star
was $\mathrm{[Fe/H]}<-3.5$, which inspired a more detailed study of this
object.

In 2008 October, we observed \object{\bd} again, using longer exposures,
a wider wavelength coverage, and higher spectral resolution. Three
different grating settings were used in order to cover 3080--9370\,
{\AA}. We employed an 0.4 arcsec slit width (and no on-chip binning),
achieving a resolving power of $R \sim 90,000$ . The total exposure time
was 120 minutes for the near-UV setting, while 20 and 25 minute exposure
times were used for the redder grating settings. We also observed a
rapidly rotating B-type star (\object{HD 12534}) with the same settings
as a reference for the elimination of telluric features. In order to
obtain a spectrum over the wavelength range 3900--3960\,{\AA}, which
fell in the CCD gap in the October run, an additional observation was
carried out in 2008 November with a setting that covers 3540--5240\,
{\AA}, using an exposure time of 5 minutes, an 0.6 arcsec slit (and
$2\times2$ on-chip binning), with a resulting resolving power of $R
\sim 60,000$. Details of the observations are provided in
Table~\ref{obslog}.

Since the data from the October run are the highest quality, we mainly
use them in this study. The November data are used for the analysis of
some atomic lines at 3900--3960\,{\AA}, and for measurement of the
radial velocity. A spectral atlas is provided as online material (see
Appendix). The August data are used only for a radial-velocity
measurement.

We carried out data reduction with the IRAF\footnote{IRAF is
distributed by the National Optical Astronomy Observatory, which is
operated by the Association of Universities for Research in Astronomy
(AURA) under cooperative agreement with the National Science
Foundation.} echelle package, performing overscan subtraction, CCD
linearity correction \citep{Tajitsu10}, cosmic-ray removal (see \S2.2 in
\citealt{Aoki05}), division by the normalized flat-field exposures, extraction
of the echelle orders, wavelength calibration, continuum
normalization, and combining of all overlapping orders. The data in
pixels affected by bad columns were removed.

The signal-to-noise ratio (S/N) per 0.9~{\kms} pixel is estimated as the
square root of the photon counts. For the October spectrum, a maximum
S/N of 550/1 is achieved at 4700\,{\AA}, where the two grating settings overlap.
The S/N decreases at shorter wavelengths to 300/1 at 3700\,{\AA}, and
to 100/1 at 3100\,{\AA}. At 4800-7000\,{\AA}, a S/N of 200--300 is
achieved. In the near-IR range ($>7000$\,{\AA}), the effective S/N
declines to $\sim100$, due to fringing on the CCD.  Note that these S/N
values should be multiplied by a factor of $\sim1.9$ in order to obtain
the S/N per resolution element, which is covered with $\sim3.5$ pixels
for a resolving power of $R \sim $90,000.

The August spectrum has a S/N per 1.8~{\kms} pixel of 300/1 redder than
$5500$\,{\AA}, decreasing to 200/1 at 4350\,{\AA}, and 150/1 at 4100\,{\AA}. For the
November spectrum, the S/N is 200/1 redder than $4900$\,{\AA}, 130/1 at
3900\,-\,3960\,{\AA}, and 70/1 at 3570\,{\AA}. With a resolving power
of $R\sim$ 60,000 (and $2\times2$ on-chip binning), the S/N per resolution
element is $\sim1.6$ times larger than the S/N per pixel.

\subsection{Line Measurements}

We are in the process of compiling a large master line list of
data from used by various studies (Aoki et al., in
preparation); this line data is complemented in the near-UV region by the line
lists employed in several other abundance studies of extremely metal-poor
stars \citep{Frebel07b,Honda06,Cohen08,Lai08}.  We make use of this list
for our present analysis.

The equivalent widths of unblended lines, measured by a Gaussian
fitting procedure, are employed to determine chemical abundances as
described in \S4. The atomic data and measured equivalent widths are listed in
Table~\ref{tab:lines}.  The error of the equivalent width measurement is
estimated as $\sigma_\mathrm{EW} = w n_\mathrm{pix}^{1/2} /
(\mathrm{S/N}) $, where $w$ is the pixel width ({\AA}),
$n_\mathrm{pix}$ is the number of pixels across the line, and S/N is
the signal-to-noise ratio per pixel \citep{Norris01}. Applying this
formula, the errors of our measurements are estimated to be smaller
than $0.1$\,m{\AA} for most of the measured lines, and at most
$0.4$\,m{\AA} for some individual cases.

For several undetected lines, we list upper limits on the equivalent
widths in Table~\ref{tab:lines}, estimated with a spectrum synthesis
approach. We also employ the spectrum synthesis technique for blended
lines and molecular features. Details are described in \S4.

\subsection{Radial Velocity}

Heliocentric radial velocities were measured for the high-resolution
spectra obtained at four different epochs: 2008 August 22, 2008 October 4,
2008 October 5, and 2008 November 16. The spectrum obtained with the reddest
setting in October 4 is not used, because the number of detected lines
that can be employed for the measurement is too small in that
wavelength range. We use 16 isolated \ion{Fe}{1} lines in the
wavelength regions 4030--4350\,{\AA} and 4420--4780\,{\AA}, which are
covered by all of the spectra for the four epochs, in order to measure
the radial velocities. The observed velocities are corrected for the motion of
the Earth, then converted to heliocentric radial velocities. The
resultant radial velocities and HJD (Heliocentric Julian Date) are
listed in Table~\ref{tab:velocity}.

Random errors in the radial velocity measurements are estimated as
$\sigma = s N^{-1/2} $, where $s$ is the sample standard deviation and
$N$ is the number of lines used ($N=16$). The derived random error is
typically $0.03\,\mathrm{km\,s^{-1}}$. Therefore, the total error is
dominated by the systematic error due to the instability of the
instrument, which is estimated to be about $0.5\,\mathrm{km\,s^{-1}}$. 

\begin{figure}
\epsscale{.70}
%\plotone{RV2.eps}
\plotone{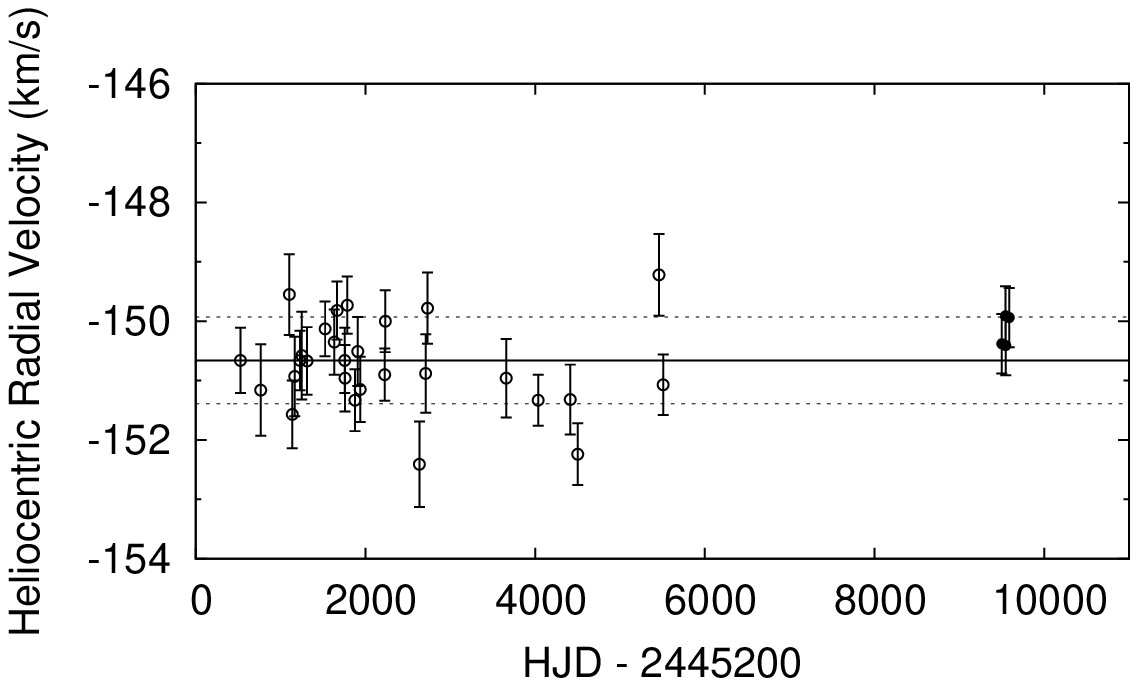}
\caption{Heliocentric radial velocity of {\bd} from 1984 to 2008.  Our
  measurements are represented by the filled circles.  The open circles
  indicate the results of \cite{Carney03}. The solid line indicates
  the mean of the entire sample ($-150.66$ {\kms}) and the dashed lines
  show the rms variation (0.73 {\kms}).}
\label{RV}
\end{figure}

We find no clear variation of the heliocentric radial velocity for
{\bd} from 2008 August 22 to 2008 November 16. Furthermore, our
measurements agree (within the reported errors) with the results of
the extended radial velocity monitoring of {\bd} from 1984 to 1997
reported by \cite{Carney03}; the results of both \cite{Carney03} and
our measurements are shown in Figure~\ref{RV}.  \cite{Carney03}
  investigated the possibility of radial velocity variations for their
  sample, including {\bd}. The ($E/I$) of {\bd} is close to unity,
  indicating that the scatter of the radial velocity (the ``rms
  external error'' $E$) is mostly explained by mean internal error of
  measurements ($I$).  The $P(\chi^{2})$ value (0.23) for {\bd}, which is
  significantly larger than the values of known binary stars, also indicates that the radial velocity scatter is explained by internal errors. Hence, though no useful data exist between their measurements and ours, no 
  signature of radial velocity variation is found for {\bd} over 24 years
(rms = 0.73 {\kms}), indicating that it is either not a
binary star, or it is one with an extremely long period.
 
\subsection{Interstellar Absorption}

\begin{figure}
\epsscale{.70}
%\plotone{NaD.eps}
\plotone{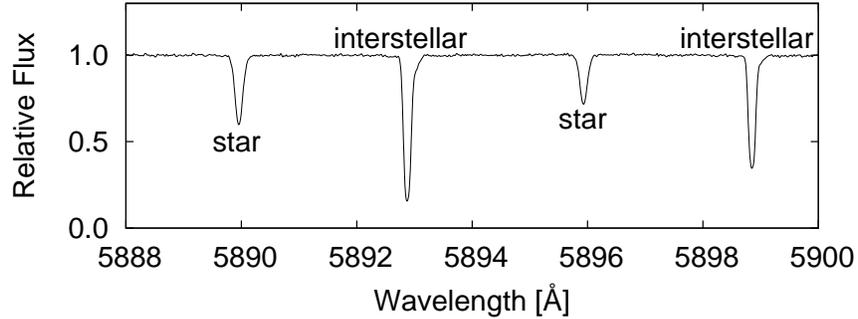}
\caption{Stellar and interstellar \ion{Na}{1} D lines in the spectrum of {\bd}.}
\label{interstellarNa}
\end{figure}

As shown in Figure~\ref{interstellarNa}, we detect a clear single
component of interstellar absorption associated with both the individual
\ion{Na}{1} D1 and D2 lines. The observed wavelengths, heliocentric
radial velocities, and equivalent widths, which were measured by fitting
Gaussian profiles, are presented in Table~\ref{interstellarD}. We
confirm that the equivalent widths measured by direct integration 
differ by less than 5\% from those obtained by Gaussian fitting.

According to the relationship between the equivalent widths of
the interstellar \ion{Na}{1} D2 line and distance shown in Figure
7 of \cite{Hobbs1974}, a lower limit on the distance to {\bd} is
estimated to be about 200\,pc. This result is consistent with the
distance derived from the stars geometric parallax (205\,pc; see \S2.5).

The color excess for {\bd} is estimated to be $E(B\!-\!V) = 0.042$, based
on empirical relations between the equivalent width of
the interstellar \ion{Na}{1} D2 line and $E(B\!-\!V)$ provided in Table 2
of \cite{Munari997}.

\subsection{Parallax, Proper Motion, and Kinematics}

The parallax and proper motion of {\bd} were measured by the Hipparcos
mission. We adopt the results of the new reduction\footnote{VizieR
  Online Data Catalog: I/311} \citep{vanLeeuwen2007}, which are listed
in Table~\ref{hipparcos}. The parallax of $4.88\pm1.06$ mas indicates
the distance of $205\,^{+57}_{-37}$\,pc.

The orbital parameters of this star (Table~\ref{tab:kinematics}) are calculated for the
above data and the radial velocity of {\bd}, following the procedures of
\citet{Carollo10}. The Galactocentic rotational velocity, $V_\phi$, of
{\bd}, its derived $Z_{\rm max}$ (the maximum distance from the
Galactic plane reached during its orbit), and its orbital
eccentricity, $e$, are typical for values associated with objects belonging to
the inner-halo population, even though most stars with such extremely
low metallicity are usually members of the outer-halo population
\citep{Carollo10}. The kinematic separation into these
populations is, however, meaningful only in a statistical sense, and is not
definitive for individual objects. Indeed, the outer-halo population has
a large dispersion in its orbital parameters, and this star may just be
in the tail of the distribution. 

\subsection{Photometry Data}

We present available photometry for {\bd} in Table~\ref{photometry}. 
Values for the $B_T$ and $V_T$ photometry are taken from the Tycho-2
Catalog\footnote{VizieR Online Data Catalog: I/259} \citep{Hog00}.
Empirical relations between the Johnson $BV$ magnitudes and
Hipparcos-Tycho $B_TV_T$ magnitudes are presented in
\cite{Bessell2000}. The near-IR $JHK_s$ magnitudes are from the 2MASS\footnote{The
Two Micron All Sky Survey (2MASS) is a joint project of the University
of Massachusetts and the Infrared Processing and Analysis
Center/California Institute of Technology, funded by the National
Aeronautics and Space Administration and the National Science
Foundation. \citep{Skrutskie06}} All-Sky Catalog of Point
Sources\footnote{VizieR Online Data Catalog: II/246} \citep{Cutri03},
while the value of the $b\!-\!y$ color and Balmer discontinuity index
$c_1 = (u\!-\!v)-(\!v-\!b)$ are from the uvby$\beta$
Catalog\footnote{VizieR Online Data Catalog: II/215} \citep{Hauck998}.
Note that the errors on $b\!-\!y\,,\,c_1$ listed in
Table~\ref{photometry} are random errors, and do not include systematic
errors.

The reddening maps of \cite{Schlegel98} and \cite{Burstein982} yield
$E(B\!-\!V) = 0.092$ and $E(B\!-\!V) = 0.115$, respectively, along the
line of sight toward \object{\bd}. However, taking the short distance to
{\bd} into account, these estimates represent upper limits to the true
reddening. Following \cite{Anthony-Twarog994}, we reduced these
estimates by the fraction $f = 1- \exp (- H r \sin |b|)$, where $r$ is
the star's distance, $b$ is the Galactic latitude, and $H$ is a constant
defined by \cite{Bond1980} as 0.008\,pc$^{-1}$. Assigning $r=205$\,pc,
as described in \S2.5, we obtain $f=0.34$. The resultant reddening
estimates are $E(B\!-\!V) = 0.031$ from the \cite{Schlegel98} maps and
$E(B\!-\!V) = 0.039$ from the \cite{Burstein982} maps.

Together with the estimate from the interstellar absorption features in
the high-resolution spectrum of {\bd} presented in \S2.4 (0.042), we
adopt a straight average of these three estimates: $E(B\!-\!V) = 0.037$.
Reddening corrections for the other passbands are derived from
$E(B\!-\!V)$ based on the relative extinctions given in Table 6 of
\cite{Schlegel98}.

\section{Stellar Parameters}\label{sec:param}

\subsection{Effective Temperature}

To estimate the effective temperatures for {\bd} from dereddened color
indices, we employ the procedures and temperature scales of
\cite{Alonso96,Alonso99, Alonso01}, \cite{Gonzalez009}, and
\cite{Casagrande10} \cite{Alonso96, Alonso99,Alonso01} and
\cite{Gonzalez009} provide the relations between {\Teff} and colors for
both dwarfs and giants. Since {\bd} is a subgiant, we calculate
temperatures for both cases. The \cite{Casagrande10} scale applies to
both dwarfs and subgiants.

\cite{Alonso96} established the temperature scale for dwarfs, while
\cite{Alonso99,Alonso01} did so for giants. However, \cite{Ryan99} pointed out that
the temperature scales for dwarfs provided by \cite{Alonso96} exhibit an
unphysical metallicity dependence at low metallicity. Since the effect
appears at $\mathrm{[Fe/H]}<-3.0$ \citep[see Figure 5 in ][]{Ryan99}, we
assume $\mathrm{[Fe/H]}=-3.0$ for {\bd} in the calculations both for the
dwarf and the giant case; this assumption does not
significantly affect the resultant temperatures.

The scales of \cite{Gonzalez009} are available over the metallicity
range $-3.5 \leq \mathrm{[Fe/H]} \leq +0.5$ for dwarfs, and over $-4.0
\leq \mathrm{[Fe/H]} \leq +0.2$ for giants. We assume
$\mathrm{[Fe/H]}=-3.5$ for the dwarf case, and $\mathrm{[Fe/H]}=-3.8$
for the giant case.

For the \cite{Casagrande10} scales, we adopt $\mathrm{[Fe/H]}=-3.7$ for
$b\!-\!y$, considering that the relation for the color is available in
the range $-3.7 \leq \mathrm{[Fe/H]} \leq +0.5$, and $\mathrm{[Fe/H]}=-3.8$ for
the other colors.

Dereddened color indices are obtained from the photometry data and
reddening estimates described in \S2.6. The transformation of
Hipparcos-Tycho $B_T V_T$ magnitudes into Johnson $BV$ is performed
using the relations given in Table 2 from \cite{Bessell2000}. For
the \cite{Alonso96,Alonso99,Alonso01} calibrations, we transform the 2MASS
$JHK_s$ magnitudes into the TCS (Telescopio Carlos Sanchez) system
using Equation (1) in \cite{Ramirez004}. Where possible, we
dereddened the colors first and then transformed onto the
required system.

All the dereddened and transformed colors are listed in
Table~\ref{Teff}, along with the derived temperatures. The uncertainties
of the temperatures arising from the photometric errors are presented.
Note that the small error on {\Teff} derived from $b\!-\!y$ is because
the photometric error on $b\!-\!y$ does not consider systematic errors.

We rely primarily on the $V\!-\!K_s$ temperature estimated from
\cite{Casagrande10}, and adopt an effective temperature of
$5430\,\mathrm{K}$ for {\bd}. The $V\!-\!K_s$ calibration is less
sensitive to metallicity and photometric errors than is the case for
other colors. Although the $(V\!-\!K_s)_0$ color depends sensitively on
the adopted reddening, we have obtained a very reliable reddening
estimate for {\bd} (see \S2.6) from multiple methods. The uncertainty due to the
photometric errors are less than 50\,K. 
 
We note here that most of the other estimates listed in
Table~\ref{Teff} are within $100\,\mathrm{K}$ of our adopted
value. Exceptions are the values determined from $B-V$ and
  $J-H$. The {\Teff} scales from $B-V$ are not well-determined,
  partially because this color index is sensitive to metallicity. The
  $J-H$ color may not be a good indicator because the wavelength
  difference between the two bands is small, and the color index is
  less sensitive to {\Teff}.

Effective temperature can also be determined by the constraint that the derived
  abundances of Fe from individual lines show no dependence on
  their excitation potentials. We investigate this in 
\S~\ref{sec:felines}. 

Another constraint on {\Teff} is obtained from the profiles of Balmer
lines, which are dependent on the temperature of the line-forming
layers of stellar atmospheres, as well as on the surface gravity. The
Balmer-line profiles calculated by
\citet{Barklem02}\footnote{http://www.astro.uu.se/~barklem/} are
compared with those of the observed spectrum. We give priority to the
H$\beta$ line, because this line is not as sensitive to non-LTE
effects as H$\alpha$ \citep{Barklem07}, and the line is almost free
from contamination from other spectral lines. Among the synthetic
spectra calculated for {\Teff}$=$5200--5600~K, {\logg}$=$3.5--3.8 and
[Fe/H]$=-3.5$, the spectrum with {\Teff}=5400~K shows the best
agreement with the H$_{\beta}$ profile of {\bd} in the wing regions
(1--10~{\AA} from the line center). The largest uncertainty is due to
the continuum placement, which is accomplished in the manner described
by \citet{Barklem02}, resulting in an uncertainty of $\pm
100$~K. Hence, we conclude {\Teff}$=5400\pm100$~K for this star. To be
conservative, we use $\Delta T_\mathrm{eff} = 150\, \mathrm{K}$ for
obtaining estimates of the abundance uncertainties in \S4.

\subsection{Microturbulence}

\begin{figure}
\epsscale{.80}
%\plotone{vmicroFeI.eps}
%\plotone{vmicroTiII.eps}
%\plotone{vmicroNiI.eps}
\plotone{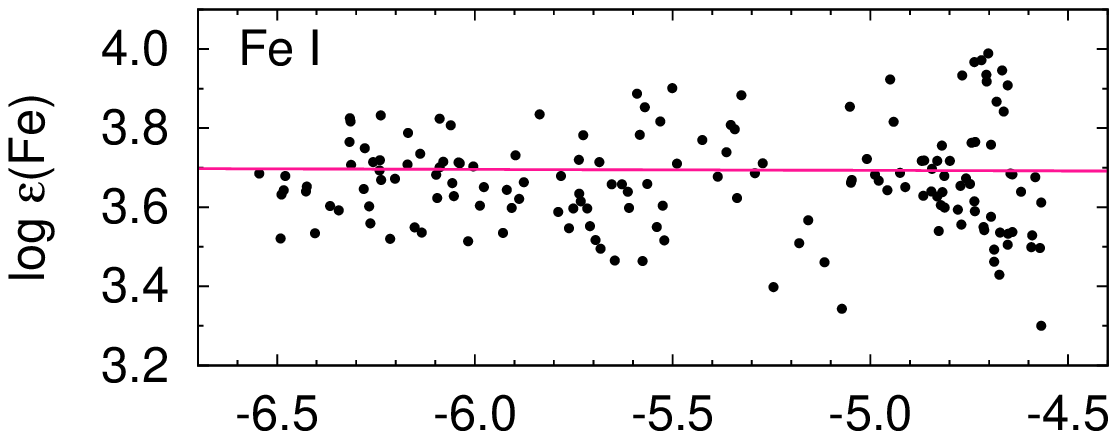}
\plotone{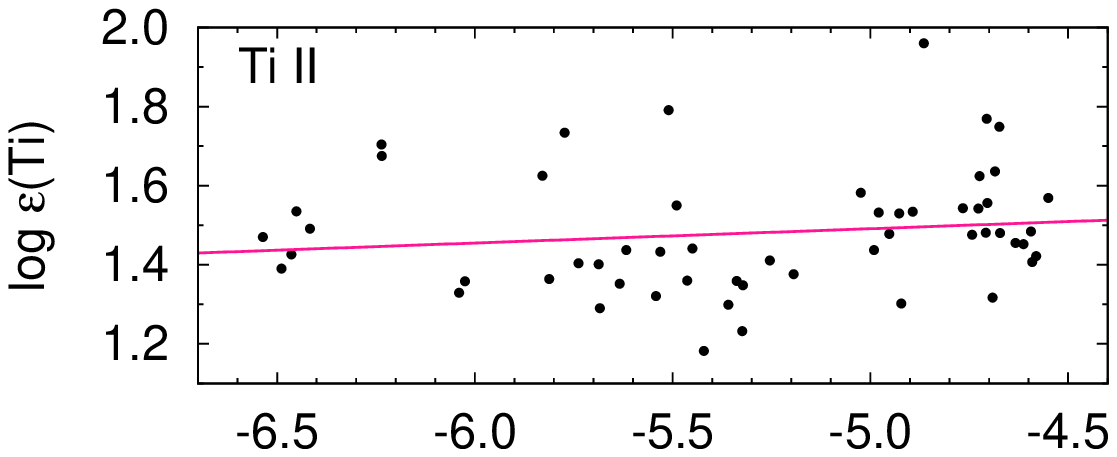}
\plotone{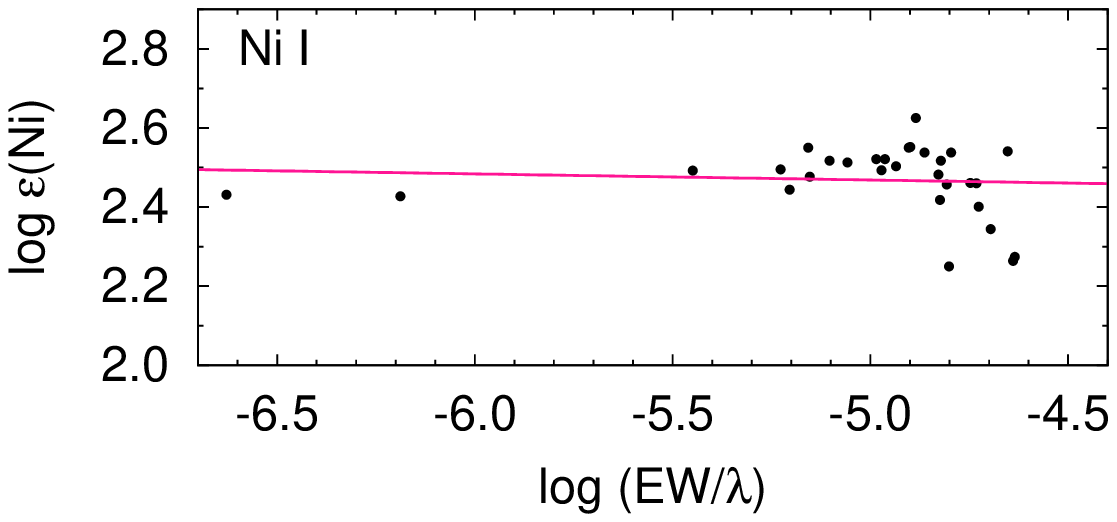}
\caption{Derived Fe, Ti, and Ni abundances, as a function of the strength of each line. The solid lines indicate the fitted linear functions.}
\label{vmicro}
\end{figure}

The microturbulent velocity, $v_\mathrm{micro}$, is derived from 158
\ion{Fe}{1} lines by demanding that no trend is found for Fe
abundances with line strengths, $\log (\mathrm{EW}/\lambda)$. In the
upper panel of Figure~\ref{vmicro}, we illustrate the situation with
the adopted value, $v_\mathrm{micro} = 1.3\,\mathrm{km\,s^{-1}}$. The
above effective temperature ({\Teff} = 5430~K) and {\logg} = 3.4 (see
below) are adopted in the calculations. No significant trend is
found for 55 \ion{Ti}{2} lines and 30 \ion{Ni}{1} lines, as indicated in
the middle and lower panels of Figure~\ref{vmicro}. The uncertainty of
the microturbulence is estimated to be $0.3\,\mathrm{km\,s^{-1}}$.

\subsection{Surface Gravity}

Surface gravity is determined from the LTE ionization equilibrium of
the Fe and Ti abundances, i.e., by requiring that the Fe and Ti abundances
derived from neutral (\ion{Fe}{1}, \ion{Ti}{1}) and singly-ionized
(\ion{Fe}{2}, \ion{Ti}{2}) lines be identical to within 0.02\,dex,
respectively. The ionization balance of both
\ion{Fe}{1}\,-\,\ion{Fe}{2} and \ion{Ti}{1}\,-\,\ion{Ti}{2} is
achieved simultaneously, when assuming $\log g=3.4$ in cgs units. We estimate
an uncertainty of 0.3\,dex in {\logg}, corresponding to
uncertainties in the abundance determinations of $\sim0.1$\,dex.

Considering how non-LTE (NLTE) effects may change the result of Fe abundances, only
\ion{Fe}{1} lines are expected to suffer from significant NLTE
corrections, while \ion{Fe}{2} lines are almost immune to them
(see Asplund et al. 2005 for a review). Although a consensus on the
expected magnitude of the NLTE effects has not been reached, NLTE
corrections for \ion{Fe}{1} estimated by previous studies are about
$+0.2$\,dex at low metallicity. If we apply an NLTE correction for
\ion{Fe}{1} of $+0.2$\,dex, the surface gravity needs to be increased by
about $+0.4$\,dex, resulting in $\log g=3.8$. Assuming this value, however,
the Ti abundances from \ion{Ti}{1} and \ion{Ti}{2} lines result in a
0.1\,dex disagreement.

We can also estimate the surface gravity from the parallax of {\bd} (see
\S2.5), employing the fundamental relation:

\begin{equation}
\log \frac{g}{g_\odot} = \log \frac{M}{M_\odot} + 4 \log
\frac{T_\mathrm{eff}}{T_{\mathrm{eff}\odot}} + 0.4 (V_0 + BC + 5 +
5 \log \pi \mathrm{[arcsec]} - M_{{\rm bol},\odot}) ,
\end{equation}

\noindent
 where $M_{{\rm bol},\odot}$ is the absolute bolometric magnitude of
  the Sun (4.75). The stellar mass $M$ is assumed to be
  $0.8\,M_\odot$. The bolometric correction ($BC$) is derived using
  Equation (18) of \cite{Alonso99}, assuming $\mathrm{[Fe/H]}=-3.0$;
  the result is $-0.256$\,mag. With our adopted temperature of
  $T_\mathrm{eff}=5430\,\mathrm{K}$, the measured parallax for {\bd}
  of $4.88\pm1.06$\,mas yields a $\log g = 3.3\pm0.2$, where the error
  of the parallax is the dominant error source. This agrees with our
  adopted value, $\log g = 3.4\pm0.3$, within the errors.

\begin{figure}
\epsscale{.60}
%\plotone{YYiso.eps}
\plotone{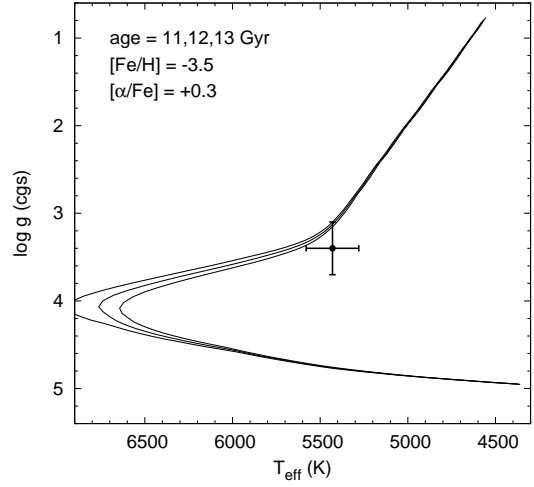}
\caption{Adopted surface gravity and effective temperature for {\bd}, with 
error bars indicating their uncertainties, in comparison with 
Yonsei-Yale \citep{Kim02,Demarque04}  
isochrones with $\mathrm{[Fe/H]}=-3.5$ and $\mathrm{[\alpha/Fe]}=+0.3$ for 11~Gyr (left), 12~Gyr (middle) and 13~Gyr (right).}.
\label{YYiso}
\end{figure}

Figure~\ref{YYiso} shows our adopted effective temperature and surface
gravity, with error bars of $\Delta T_\mathrm{eff} = 150\,
\mathrm{K}$ and $\Delta \log g = 0.3\,\mathrm{dex}$, compared with
the Yonsei-Yale isochrone \citep{Kim02,Demarque04}, computed for an age
of 12\,Gyr, metallicity of $\mathrm{[Fe/H]}=-3.5$, and a 0.3\,dex
$\alpha$-element enhancement ($\mathrm{[\alpha/Fe]}=+0.3$). The isochrone
indicates two possibilities for the derived effective temperature of
{\bd} ($T_\mathrm{eff}=5430\, \mathrm{K}$): a subgiant case ($\log g
\sim 3.1$) and a dwarf case ($\log g \sim 4.8$). Our LTE result is
consistent with the subgiant case, while the dwarf case is clearly
excluded. As described above, the NLTE correction for \ion{Fe}{1}
results in a higher surface gravity ($\log g \sim 3.8$), which
corresponds to an unrealistic result, lying between the dwarf and
subgiant cases, in the isochrone. This suggests that, in reality, the
NLTE correction for \ion{Fe}{1} is not as large as 0.2\,dex.

\subsection{Iron Abundance}\label{sec:felines}

We derived Fe abundances from 158 \ion{Fe}{1} and 11 \ion{Fe}{2}
lines, following the descriptions in \S4. Figure~\ref{Fe-lambda} shows
Fe abundances determined from individual \ion{Fe}{1} and \ion{Fe}{2}
lines, as a function of wavelength. The wide wavelength coverage
and high quality of our spectrum allows us to investigate many lines,
despite the low metallicity of this star. In particular, four
\ion{Fe}{2} lines at 3200\,-\,3300\,{\AA} with equivalent widths
of more than 15\,m{\AA} (see Table~\ref{tab:lines}) are important
for surface gravity determination from the
\ion{Fe}{1}\,/\,\ion{Fe}{2} ionization equilibrium, since most optical
\ion{Fe}{2} lines are weak.

\begin{figure}
\epsscale{.80}
%\plotone{excitation.eps}
\plotone{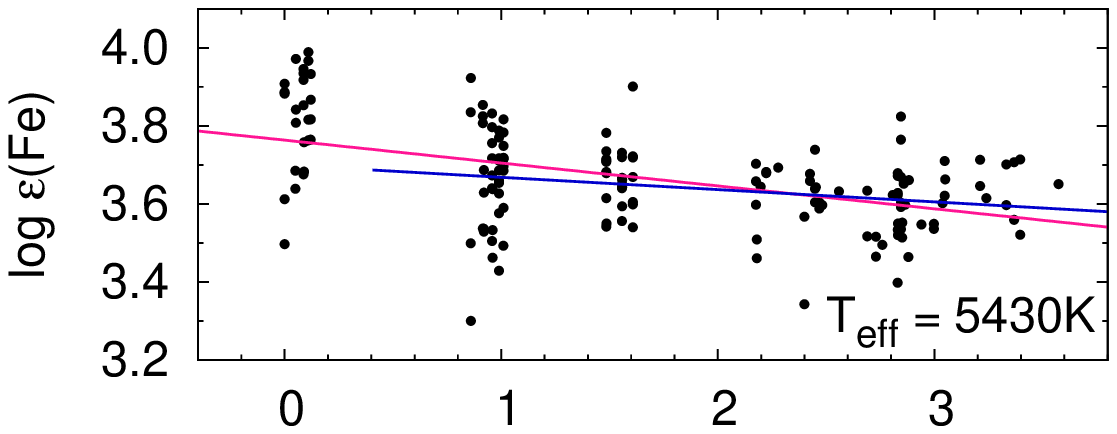}
%\includegraphics{excitation.eps}
%\plotone{excitation-100.eps}
\plotone{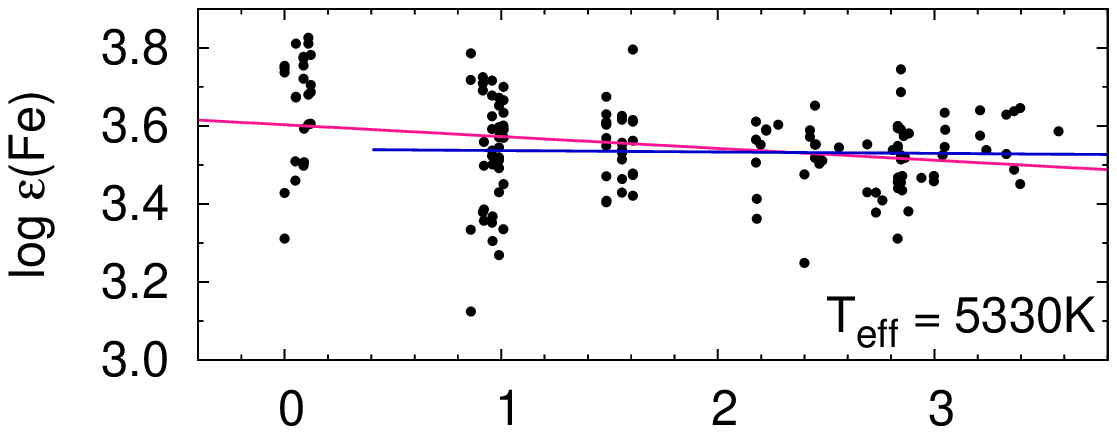}
%\includegraphics{excitation-100.eps}
%\plotone{excitation-200.eps}
\plotone{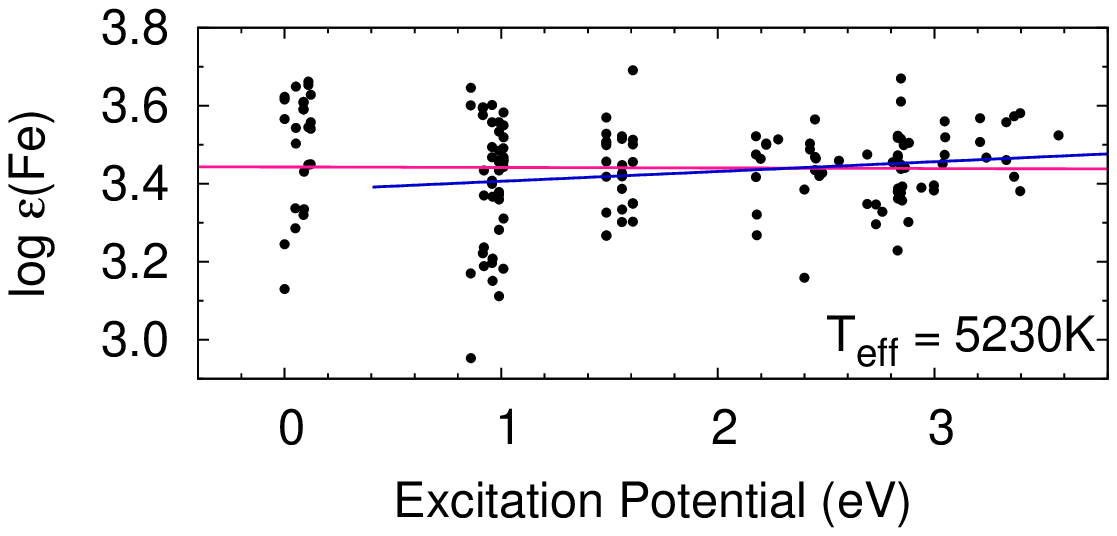}
\caption{Fe abundances determined from individual \ion{Fe}{1}
  lines, as a function of excitation potential. The upper panel is the case
  for our adopted temperature, $T_\mathrm{eff}=5430\,\mathrm{K}$,
  while the middle and lower panels apply for $T_\mathrm{eff}$ lower by 
  $100\,\mathrm{K}$ and $200\,\mathrm{K}$, respectively. The longer
  lines indicate the fitted linear functions to all the points, while
  the shorter lines apply only to the points with $\chi>0.2 \,\mathrm{eV}$.}
\label{excitation}
\end{figure}

\begin{figure}
\epsscale{.70}
%\plotone{Fe-lambda.eps}
\plotone{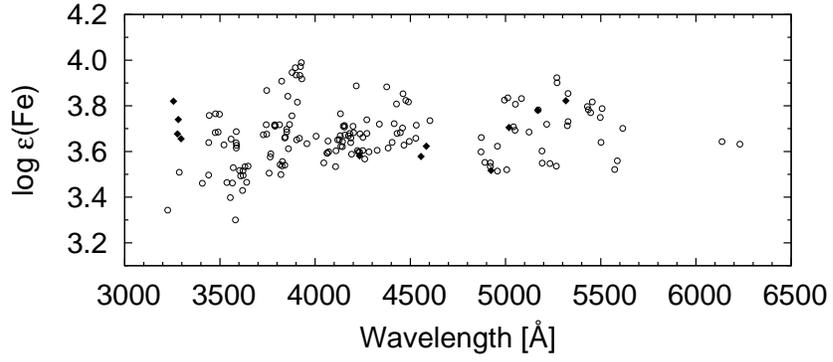}
\caption{Fe abundances determined from individual \ion{Fe}{1} and
  \ion{Fe}{2} lines, as a function of wavelength. The open circles
  and filled diamonds represent \ion{Fe}{1} and \ion{Fe}{2} lines,
  respectively.}
\label{Fe-lambda}
\end{figure}

The upper panel of Figure~\ref{excitation} shows iron abundances
determined from 158 measured \ion{Fe}{1} lines, as a function of
excitation potential. No trend is expected in such a plot, when the
excitation equilibrium is satisfied. However, as indicated by the longer
line in the upper panel of Figure~\ref{excitation}, a slope of about
$-0.06\;\mathrm{dex\;eV^{-1}}$ is found by fitting a linear function to
all the points. This trend might imply that our photometric temperature
is slightly too high. 

Some previous studies of metal-poor stars, however, have also reported
disagreement between photometrically-derived temperatures and
spectroscopically-derived temperatures that are estimated so as to
achieve excitation equilibrium \citep[e.g., ][]{Cohen08}, and speculated
that the problem arises from the use of lines with excitation potentials
of $\chi\sim0\,\mathrm{eV}$. If the fitting is performed using only the
data points in $\chi>0.2\,\mathrm{eV}$, the slope becomes much
shallower, although a weak trend remains (lines in the top panel of 
Figure~\ref{excitation}), as also found by \cite{Lai08} for their very
metal-poor stars. 

 The slope found by fitting a linear function to the 133 points with
  $\chi>0.2\,\mathrm{eV}$ is $-0.033 \pm 0.010\;\mathrm{dex\;eV^{-1}}$
  when {\Teff}$=5430$~K is adopted (the top panel of
  Figure~\ref{excitation}). The null hypothesis that there is no
  correlation between the two values is rejected by the regression
  analysis at the 99.5\% confidence level. If the abundance analysis
is made assuming {\Teff}$ = 5330\,\mathrm{K}$, no trend for the lines
with $\chi>0.2\,\mathrm{eV}$ appears (the short line in the middle
panel of Figure~\ref{excitation}). The random error is estimated to be 50~K,
adopting the 95\% confidence level in the regression analysis.

 We note for completeness that, in order to achieve the excitation
 equilibrium for all the lines, the temperature needs to be decreased
 by $200\, \mathrm{K}$ from the adopted value (see the bottom of
 Figure~\ref{excitation}).

 The {\Teff} estimated from the analysis of the Fe lines, assuming no
  dependence of Fe abundances on the excitation potential of
  individual lines, is $5330\pm50$~K if the lines with
  $\chi\sim0\,\mathrm{eV}$ are excluded. Although this is 100~K lower than the
  value estimated from the color index in \S~\ref{sec:param}, as found
  for metal-poor stars by previous studies, the difference is still
  smaller than the error adopted in our analysis.

%Figure~\ref{Fe-lambda} shows Fe abundances derived from individual
%lines, as a function of wavelength. No evident correlation is found in
%the plot.

The derived abundances are $\mathrm{[Fe/H]}=-3.83$ from \ion{Fe}{1} and
$\mathrm{[Fe/H]}=-3.82$ from \ion{Fe}{2}. In the following discussion,
we adopt the result for \ion{Fe}{1}, $\mathrm{[Fe/H]}=-3.83$, as our
final iron abundance estimate for {\bd}. Abundance errors for iron are
described below in \S4.

\subsection{Adopted Stellar Parameters}

We list our adopted stellar parameters and their uncertainties for
{\bd}, which are used for the abundance analysis, in
Table~\ref{parameters}. The adopted {\Teff} is 80~K lower than that
adopted by \citet{Ito09}, who simply adopted the value of
\citet{Carney03}. The surface gravity is lower by 0.3~dex than that of
\citet{Ito09}, partially due to the lower {\Teff} adopted by the present
work. The metallicity is also slightly lower than that of \citet{Ito09},
also due to the lower {\Teff} used in this work.  

If the lower effective temperature suggested from the analysis of
  \ion{Fe}{1} lines ({\Teff} = 5330~K, \S~\ref{sec:felines}) is
  adopted, {\logg} = 3.2 and [Fe/H] $=-3.9$ are derived. The change of
  micro-turbulent velocity is smaller than 0.1~{\kms}. The {\Teff} and
  {\logg} also agree within the errors with the subgiant branch of the
  isochrones shown in Fig.~\ref{YYiso}, although the data point lies
  slightly below the isochrones.

\section{Abundance Analysis and Results}\label{sec:ana}

\subsection{Abundance Determinations}

We adopt one-dimensional LTE model atmospheres for the stellar
parameters of {\bd}, interpolating within the grid of ATLAS9 ODFNEW model
atmospheres \citep{Kurucz1993a,Castelli003}. The computed equivalent
widths of spectral lines are compared with the observed ones listed in
Table~\ref{tab:lines}, in order to derive elemental abundances. The
microturbulence velocity determined for {\bd} in \S3 is adopted in the
spectral line-formation calculations.

For molecular features and blended lines, we calculate synthetic
spectra to compare with the observed spectrum directly, changing the
abundance by 0.05\,dex until the computed and observed spectra are
well-matched. The spectrum synthesis code that we use is based on the
same assumptions as the model atmosphere program of
\cite{Tsuji1978}. We have assumed a Gaussian profile to account for
the broadening by macroturbulence and the instrument, assuming that 
rotation is not a dominant source of the broadening. A broadening of
$6.0\,\mathrm{km\,s^{-1}}$ is adopted for the profiles of single lines.

We also employ the spectrum synthesis approach to estimate upper
limits on elemental abundances for several undetected lines. The upper
limits are computed by comparing synthetic and observed spectra, and
adjusting the abundance by 0.05\,dex until the computed strength of
the line is of the same order as the noise in the observed spectrum.

The derived abundances and upper limits are listed in
Table~\ref{abundances}. The number of lines used in the analysis
  is given in the third column. When only a single line or feature is
  available, the line or feature is given in the last column .
Solar abundances are taken from
\cite{Asplund09} to obtain the abundance ratios ([X/Fe]). Details of the
analyses for individual species are described below.

\subsection{Abundance Errors}

We estimate abundance errors, within the 1D LTE formalism, and list
the results in Table~\ref{errors}.

Random errors arising from the measurements of atomic lines are
estimated as $\sigma = s {N}^{-1/2} $, where $s$ is the sample standard
deviation and $N$ is the number of lines used in the analysis. When
the number of lines is too small to estimate a valid sample standard
deviation (in the case of $N<10$), the $s$ derived from \ion{Fe}{1} lines is
used instead. For Li and the CNO elements, for which the spectrum synthesis
approach is applied, random errors are estimated by taking into account
continuum placement uncertainties, abundance variations from different
features, and how well the synthetic spectrum fits the observed
spectrum.

Abundance errors also come from the uncertainties in the stellar
parameters. We examine them by performing abundance analyses after
changing each stellar parameter (effective temperature, surface
gravity, and microturblent velocity) by its uncertainty, as estimated
in \S3 (Table~\ref{errors}). The effect of the uncertainty in
metallicity is negligible compared with those in the other parameters.

These errors are combined following the equation (2) of
  \citet{Johnson2002} where the effects of correlations between
  parameters are included (Table~\ref{errors}). The covariances are
  calculated assuming Gaussian distribution with standard deviations
  of 150~K and 0.3~dex for {\Teff} and {\logg}, respectively. We found
  that the correlation between {\Teff} and {\logg} have small impact
  on the total error, while effects of others are negligible.  To
estimate the $\sigma_{\rm total}$ of [X/Fe], changes of the abundance
ratio ([X/Fe]) corresponding to the changes of individual parameters
are adopted in the above calculation.

Possible NLTE and 3D effects
are discussed for individual species in \S4.3\,-\,4.7.

%{\bf Teff errors need to be updated if sigma Teff is changed to be 100K.}
%The root-sum-square of these
%  errors ($\sigma_{\rm total}$ = $(\sigma_{\rm random}^{2} +
%  \sigma_{T_{\rm eff}}^{2} + \sigma_{\log g}^{2} +
%  \sigma_{v_{\rm micro}}^{2})^{-1/2}$) is adopted as the total error. 

\subsection{Light Elements: Li and Be}

\subsubsection{Lithium}

\begin{figure}
\epsscale{.70}
%\plotone{Li.eps}
\plotone{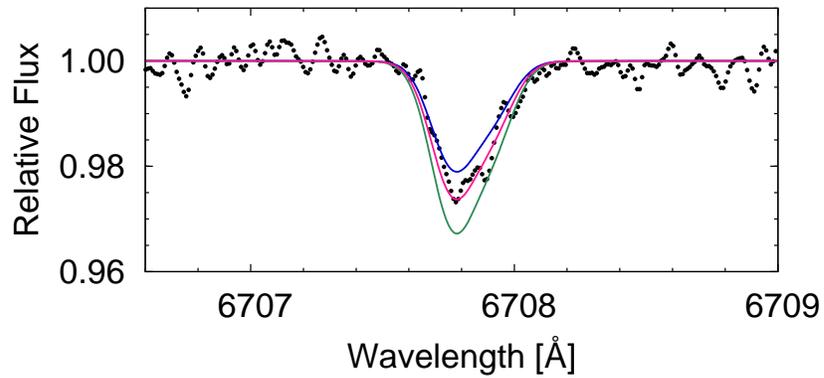}
\caption{Observed spectrum (dots) and synthetic spectra (lines) for the \ion{Li}{1} doublet at 6708\,{\AA}. Assumed abundances are $\log \epsilon(\mathrm{Li}) =+0.95$, +1.00, and +1.05.}
\label{Lithium}
\end{figure}

We have employed spectrum synthesis to measure the lithium abundance for
{\bd} from the \ion{Li}{1} doublet at 6708\,{\AA}. Atomic data for these
lines are taken from \cite{Smith98}; we assume no contribution
from $^6$Li. Figure~\ref{Lithium} shows the observed spectrum compared
with synthetic spectra of different Li abundances. The derived
result is $\log \epsilon(\mathrm{Li}) =1.0$.

There are several studies on the effects of NLTE on Li abundances derived from the
\ion{Li}{1} doublet at 6708\,{\AA}. The most recent and sophisticated
is by \cite{Lind09}, who performed NLTE calculations for a wide
range of stellar parameters. For parameter sets similar to that of
{\bd} ($T_\mathrm{eff}=5500\,\mathrm{K} ,\: \log g=[3.0, 4.0] ,\:
\mathrm{[Fe/H]}=-3.0 ,\: v_\mathrm{micro}=[1.0, 2.0] ,\: \log
\epsilon(\mathrm{Li})=[0.9,1.2]$), only very small NLTE
corrections are suggested: $\log \epsilon(\mathrm{Li})_\mathrm{NLTE} - \log
\epsilon(\mathrm{Li})_\mathrm{LTE} = [-0.02, 0.00]$. This indicates
that NLTE effects on the Li abundance of {\bd} have no significant
impact.

%NLTE calculations for Li with 3D model atmospheres have been very
%limited so far. 

\cite{Asplund03} presented 3D NLTE calculations for Li
abundances of two stars, and found that the 3D NLTE abundances were
similar to the 1D NLTE results. Since the metallicity assumed in their
calculation for the subgiant case in \cite{Asplund03} is one order of
magnitude higher than that of {\bd}, further investigation of 3D NLTE
corrections for the Li abundance is required to derive the final conclusion
for {\bd}. However, the Li in {\bd} is clearly lower than the value of
the Spite plateau (A(Li) $\sim 2.2$; e.g., \citealt{Melendez10}).

\subsubsection{Beryllium}

\begin{figure}
\epsscale{.70}
%\plotone{Be.eps}
\plotone{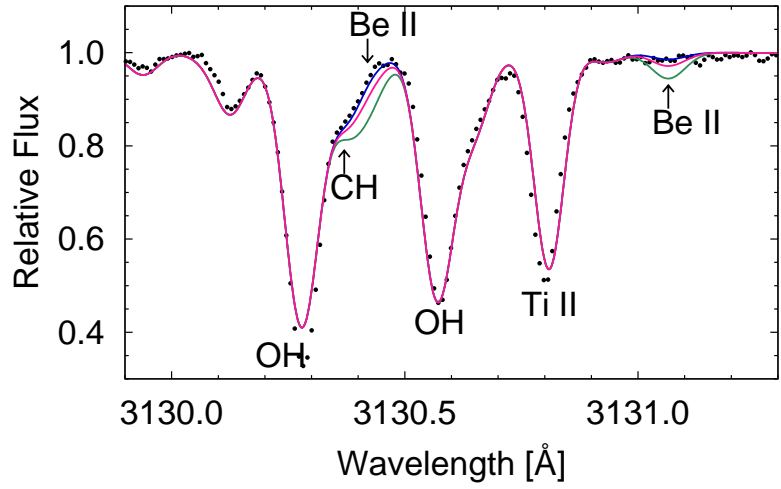}
\caption{Same as Figure~\ref{Lithium}, but for \ion{Be}{2} lines. Assumed abundances are $\log \epsilon(\mathrm{Be}) =-2.1$, $-1.8$, and $-1.5$.}
\label{Beryllium}
\end{figure}

Our high-quality near-UV spectrum of {\bd} enables inspection of the
\ion{Be}{2} lines at 3130\,{\AA}. Although the lines are not detected
in the spectrum, we set an upper limit on the Be abundance by spectrum
synthesis. The line data used is based on \cite{Boesgaard99} and
the Kurucz line
list\footnote{http://kurucz.harvard.edu/linelists.html}. We do not
include the two CH lines at 3131\,{\AA} given in Table 4 of
\cite{Boesgaard99}, because no signature of these lines are found in
our spectrum. Even if the \ion{Be}{2} line at 3131\,{\AA} is removed
in the calculations, the computed absorption strength at this
wavelength exceeds the noise level of the observed spectrum due to
these two CH lines. We note that this treatment results in a
conservative upper limit for the Be abundance. Figure~\ref{Beryllium}
compares the observed spectrum with synthetic spectra for different Be 
abundances. We estimate that the upper limit for the Be abundance is
$\log \epsilon(\mathrm{Be}) <-1.8$, and note that 3D and NLTE effects
on the Beryllium abundance are expected to be small \citep{Asplund2005}.

\subsection{C, N, and O abundances}

The abundances of the elements C, N, and O are determined from molecular
features (CH, NH, and OH) by comparison of the observed spectrum with
synthetic spectra. The abundance measurements for nitrogen and oxygen
are made by analyses of NH and OH lines in the wavelength range
$<3400$\,{\AA}.

\subsubsection{Carbon}

\begin{figure}
\epsscale{.70}
%\plotone{CH4312.eps}
%\plotone{CH4324.eps}
%\plotone{CHUV.eps}
\plotone{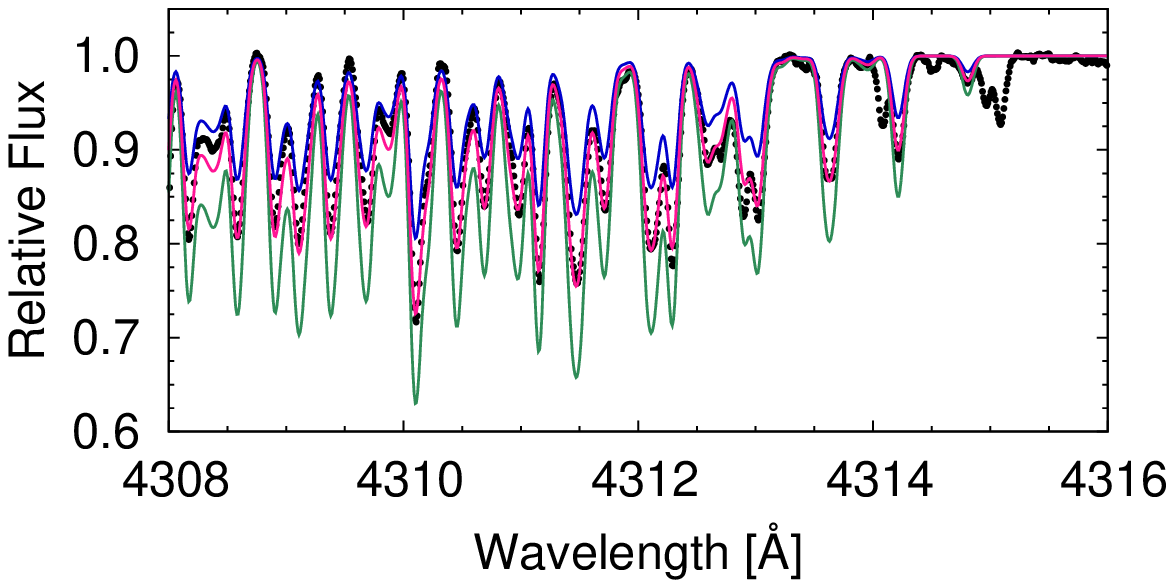}
\plotone{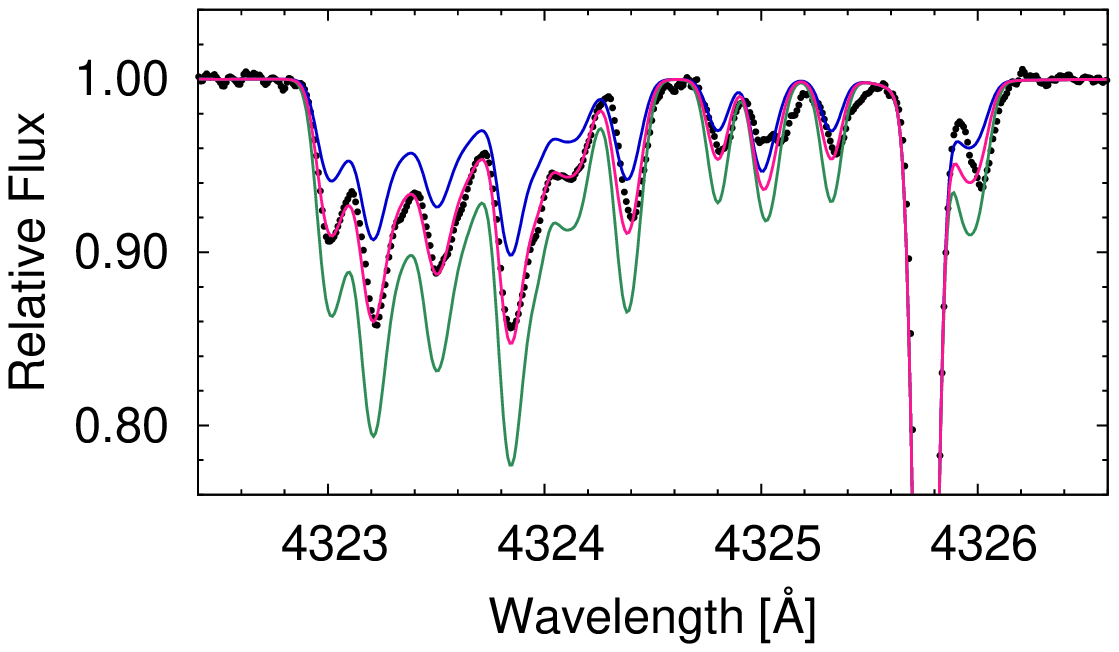}
\plotone{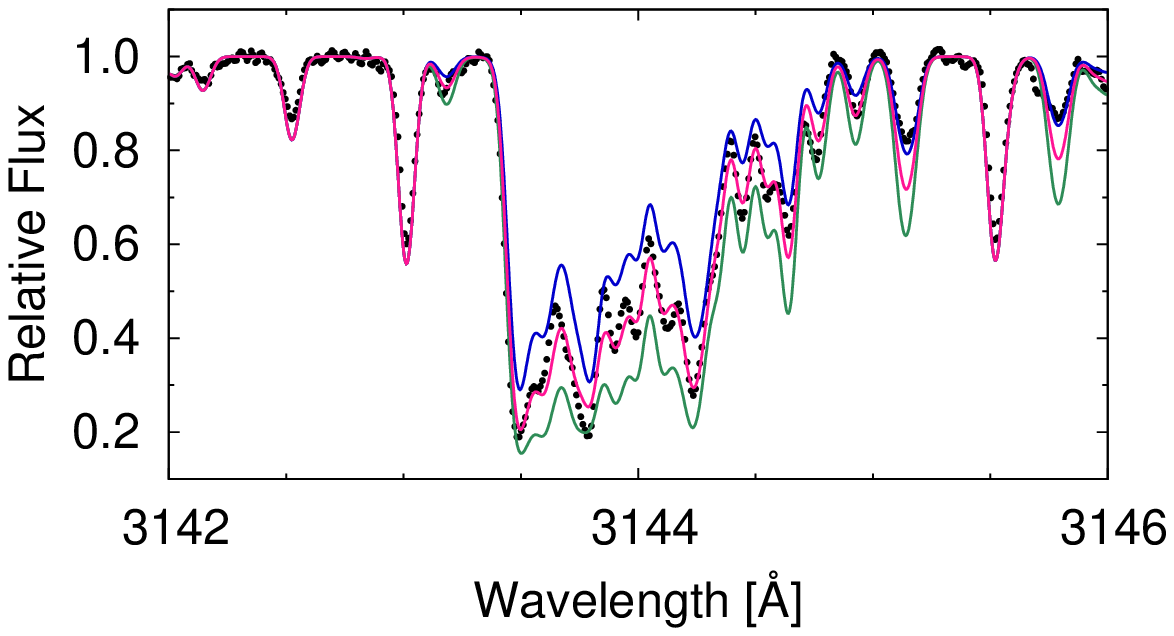}
\caption{Same as Figure~\ref{Lithium}, but for the CH features at 4312\,{\AA} (top panel), 4324\,{\AA} (middle panel) and 3144\,{\AA} (bottom panel). Assumed abundances are $\mathrm{[C/Fe]} =+1.20$, +1.40, and +1.60 for the 4312\,{\AA} feature, $\mathrm{[C/Fe]} =+1.10$, +1.30, and +1.50 for that at 4324\,{\AA}, and $\mathrm{[C/Fe]} =+1.05$, +1.25, and +1.45.}
\label{fig:CH}
\end{figure}

We measure C abundances from the CH {\it A-X} features at 4312\,{\AA}
and 4324\,{\AA} (Figure~\ref{fig:CH}). Details of the line list used
can be found in \cite{Aoki06}. The two C abundances differ by only
about 0.1\,dex. We take the straight average of them to obtain our
final C abundance, $\mathrm{[C/Fe]}=+1.35$, indicating that {\bd} has a
large excess of carbon.

We also synthesized spectra for a strong CH feature at 3144\,{\AA},
using the CH {\it C-X} line list based on \cite{Kurucz1993b}
(Figure~\ref{fig:CH}, bottom panel). Since many OH lines exist at this wavelength
region, and the accuracy of the line data is not well known, the C
abundance from this feature is not as certain as those from the
optical features. Hence, we do not include it in the final abundance
result. However, the best fit is achieved for
$\mathrm{[C/Fe]}= +1.25$, which is only 0.1\,dex below the value 
estimated from the optical features. This supports the
robustness of the C/O ratios derived by our analyses, in which the oxygen
abundance is also derived from the OH lines in the near-UV range.

\begin{figure}
\epsscale{.70}
%\plotone{CISO.eps}
\plotone{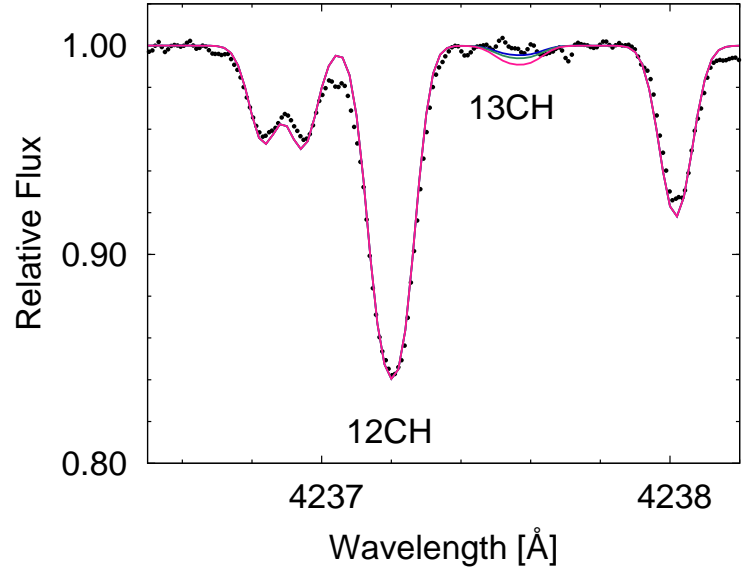}
\caption{Same as Figure~\ref{Lithium}, but for $^{12}$CH and $^{13}$CH
features at 4237\,{\AA}. Assumed $^{12}$C/$^{13}$C ratios are 20, 30, and 40.}
\label{fig:ciso}
\end{figure}

No feature of $^{13}$CH is detected in our spectrum of {\bd}. A
lower limit on the carbon isotope ratio ($^{12}$C/$^{13}$C) is
estimated from the CH lines in 4215-4240~{\AA}, as also done by
\citet{Honda04}. An example is shown in Figure~\ref{fig:ciso}. The
lower limit estimated from five features in this wavelength region is
$^{12}$C/$^{13}$C$>30$. Although the constraint is not strong, the
result suggests that $^{12}$C production by the triple-$\alpha$ reaction
dominates the contribution from the CNO cycle.

\subsubsection{Nitrogen}

\begin{figure}
\epsscale{.70}
%\plotone{NH.eps}
\plotone{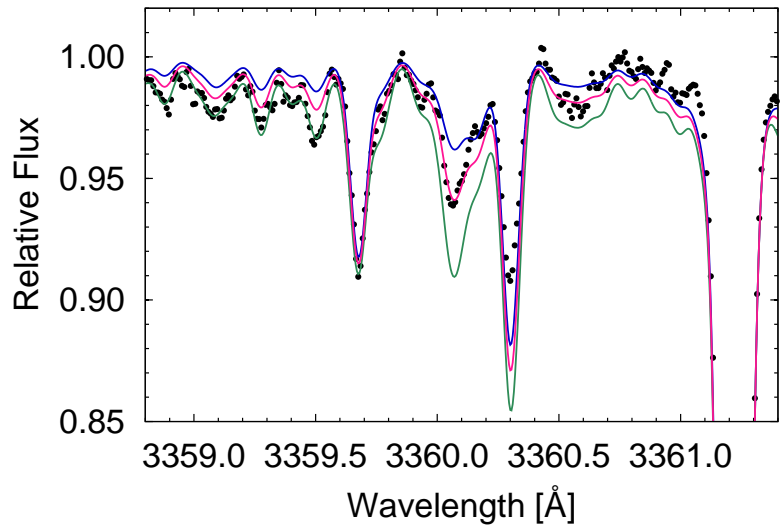}
\caption{Same as Figure~\ref{Lithium}, but for NH features at 3360\,{\AA}. Assumed abundances are $\mathrm{[N/Fe]} =+0.25$, +0.45, and +0.65.}
\label{NH}
\end{figure}

The N abundance is determined from the NH band feature at 3360\,{\AA},
using the NH line list based on \cite{Kurucz1993b} (Figure~\ref{NH}).
\cite{Aoki06} modified the $gf$-values of the \cite{Kurucz1993b} NH line list to reproduce the solar spectrum with the
solar N abundance of \cite{Asplund05}. This modification produces a
nitrogen abundance higher by +0.4\,dex. Since our adopted solar N
abundance \citep{Asplund09} is 0.05\,dex lower than given in
\cite{Asplund05}, we apply a correction of +0.35\,dex. The final result
is $\mathrm{[N/Fe]}=+0.40$, indicating that N is not significantly
over-abundant in {\bd}.

%Using the uncorrected
%line lists in the analysis, we are thus required to correct our N
%abundance.

\subsubsection{Oxygen}

\begin{figure}
\epsscale{.70}
%\plotone{OH.eps}
\plotone{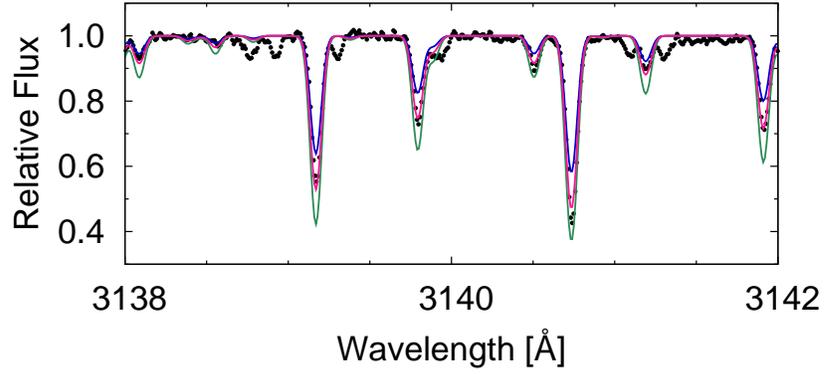}
\caption{Same as Figure~\ref{Lithium}, but for OH features at 3140\,{\AA}. Assumed abundances are $\mathrm{[O/Fe]} =+1.44$, +1.64, and +1.84.}
\label{OH}
\end{figure}

We measure 11 OH lines in the wavelength range 3120\,-\,3180\,{\AA}
(3123.95, 3127.69, 3128.29, 3139.17, 3140.73, 3145.52, 3151.00, 3166.34,
3167.17, 3172.99, 3173.20 {\AA}) to determine the O abundance, using the
line list of \cite{Kurucz1993b}. An example of spectra including OH
lines is shown in Figure~\ref{OH}. All the abundances determined from
the individual lines agree with each other to within $\pm 0.2$\,dex. A simple
average is adopted as our final result: $\mathrm{[O/Fe]}=+1.64$.
Thus, {\bd} is remarkably enhanced in oxygen.

%{\bf OH lines were measured via spectrum synthesis, but they can be
%measured by gaussian fitting and we can list EWs, which may be better.}

Neither the forbidden [\ion{O}{1}] line at 6300\,{\AA}, nor the
\ion{O}{1} triplet at 7772\,{\AA}, are detected in the observed
spectrum of {\bd}. We estimate upper limits on the O abundance from these
lines, and obtain $\mathrm{[O/Fe]}<+2.34$ for the [\ion{O}{1}]
6300\,{\AA} line, and $\mathrm{[O/Fe]}<+2.14$ for the \ion{O}{1}
7772\,{\AA} line. These upper limits are not inconsistent with the O
abundance determined from the OH lines in the UV regions.

Some previous studies discuss the discrepancies between the O
abundances determined from the three different indicators: the OH
band, the forbidden [\ion{O}{1}] line, and \ion{O}{1} lines
\citep[e.g., ][]{Israelian01,GarciaPerez06}. No constraint on this
issue is obtained for {\bd}, as only weak upper limits are obtained
from the [\ion{O}{1}] and \ion{O}{1} lines.

\subsubsection{Possible 3D and NLTE effects}

Recent calculations based on 3D hydrodynamical model atmospheres
suggest that the CH, NH, and OH bands can be severely affected by 3D
effects at low metallicity (see \citealt{Asplund2005} for a
review). The 3D models yield much lower abundances than 1D models, and
the magnitude can amount to as much as $-0.9$\,dex at
$\mathrm{[Fe/H]}=-3$ (e.g., \citealt{Asplund001}). This is because 
molecule formation in the cool regions occurs in the upper layers of
the atmospheres in the 3D case. Although NLTE calculations result in
higher abundances, they do not appear to fully compensate for the 3D
effects. Thus, we might overestimate the CNO abundances in our 1D LTE
analysis. However, it should be noted that the 3D and NLTE effects are
expected to be more or less similar for the CH, NH, and OH bands, and
abundance ratios among the three elements (e.g., C/O) are
comparatively robust. We also note that carbon and oxygen abundances
in EMP stars have been determined from CH and OH molecular features, as
done for {\bd} in the present work. Hence, the excesses of these
elements for this star are clearly evident.

\subsection{Even-Z Elements: Mg, Si, S, Ca, and Ti}

The Mg abundance is determined from eight \ion{Mg}{1} lines in the
wavelenth range 3800\,-\,5600\,{\AA} ($\mathrm{[Mg/Fe]}=+0.46$). Agreement of
the results from the eight lines is fairly good. Recently,
\cite{Andrievsky10} performed NLTE calculations for \ion{Mg}{1} lines
for a large sample of metal-poor stars, and derived 0.3\,dex higher Mg
abundances for both giants and dwarfs at $\mathrm{[Fe/H]}\sim-3$. Our
Mg abundance will also be increased if NLTE effects are taken into
account.

Since the \ion{Si}{1} line at 3906\,{\AA} falls in the CCD gap in the
October spectrum, we use the November spectrum for the Si abundance
measurement. Although the quality of that spectrum is not as high as
that of the October spectrum (as stated in \S2), it is sufficient for
a clear detection of this line. We derive a Si abundance ratio
$\mathrm{[Si/Fe]}=+0.49$. The blend with the CH lines is not severe at this
wavelength, and Gaussian fitting to the \ion{Si}{1} line works well. The
NLTE effect on the line is expected to be small ($<0.1$\,dex) for the
effective temperature of {\bd} \citep{Shi09}.

We have searched for the strongest \ion{S}{1} lines at 9213\,{\AA} and
9237\,{\AA} in the wavelength range of our spectrum
(3080\,-\,9370\,{\AA}), but find no feature at 9213\,{\AA}, and
unfortunately, the 9237\,{\AA} line falls in the echelle order
gap. The estimated upper limit on the S abundance from the 9213\,{\AA}
line is $\mathrm{[S/Fe]}<+1.06$. According to \cite{Takeda05}, the
NLTE correction for the line varies greatly with stellar
parameters. For {\bd} it may reach about $-0.3$\,dex.

In addition to the \ion{Ca}{1} lines in the wavelength range 4220\,-\,
6440\,{\AA}, we employed the \ion{Ca}{2} line at 3181\,{\AA} and the
strong \ion{Ca}{2} triplet at 8490\,-\,8670\,{\AA} to derive the Ca
abundance. These three indicators yield different results,
$\mathrm{[Ca/Fe]}=+0.31$, +0.47, and +0.91, respectively. In particular,
the large discrepancy between the Ca abundance obtained from the
\ion{Ca}{1} lines and the
\ion{Ca}{2} triplet cannot be explained by random errors (see
Table~\ref{errors}). The NLTE calculations conducted by
\citet{Mashonkina07} result in increases of the abundance from
\ion{Ca}{1} lines by more than 0.2\,dex, while 
the abundances estimated from the \ion{Ca}{2} triplet decrease by about
0.3\,dex, for similar parameter sets as {\bd}. We adopt the \ion{Ca}{1}
lines as the abundance indicator in this paper in order to compare it with
previous studies, most of which have adopted \ion{Ca}{1} lines for their
analyses. 

% our final result is $\mathrm{[Ca/Fe]}=+0.31$, because most
%previous studies of metal-poor stars derived Ca abundances from
%\ion{Ca}{1} lines, and they are favorable for comparison purposes.

We detect a number of \ion{Ti}{2} lines, and several weak \ion{Ti}{1}
lines, in the observed spectrum. The ionization equilibrium is
for this element is satisfied by our {\logg} (see \S3.3). The derived Ti
abundance ratio is $\mathrm{[Ti/Fe]}=+0.36$. Unfortunately, we could
find no NLTE study on Ti for extremely metal-poor stars discussed in the literature.

The abundances of the $\alpha$-elements Mg, Si, Ca, and Ti are
enhanced by $\sim$\,0.3\,-\,0.5\,dex with respect to Fe, as found for
most of very and extremely metal-poor stars of the Milky Way halo.

\subsection{Odd-Z Elements: Na, Al, K, and Sc}

We determine the Na abundance ratio from the \ion{Na}{1} D resonance lines
at 5890\,{\AA}, and obtain
$\mathrm{[Na/Fe]}=+0.30$. \cite{Andrievsky07} suggested an NLTE
correction of about $-0.1$\,dex on the D lines for similar stellar
parameters as {\bd}.

The only available line that could be used for an Al abundance
determination is the 
\ion{Al}{1} resonance line at 3961\,{\AA}, yielding
$\mathrm{[Al/Fe]}=-0.56$. We do not use the \ion{Al}{1} line at 3944\,
{\AA}, due to blending by CH lines. The Al abundance is quite 
sensitive to NLTE effects. From inspection of Figure 2 in
\cite{Andrievsky08}, the NLTE correction for {\bd} probably exceeds
$+0.5$\,dex, and may amount to as much as $+0.7$\,dex. Hence, 
Al is not as under-abundant as found by the LTE analysis.

The K abundance can be estimated from the \ion{K}{1} doublet at
7665\,{\AA} and 7699\,{\AA}. However, in the observed spectrum of
{\bd}, the stronger feature at 7665\,{\AA} is unfortunately superimposed
on large telluric lines. Although we attempted to remove the telluric
features, we were not successful for these strong lines. We inspect the
spectrum at 7699\,{\AA}, and set an upper limit on the K abundance of
$\mathrm{[K/Fe]}<+0.75$. According to recent NLTE calculations, K
abundances are overestimated in LTE computations. \cite{Takeda09} and
\cite{Andrievsky10} derived NLTE K abundances for the same sample of
metal-poor giants, which was originally studied under LTE by
\citet{Cayrel04}, and reached similar conclusions. According to their
estimates, the NLTE effect in EMP stars is not as large as in less
metal-poor stars.

% becomes smaller at lower metallicity; the NLTE
%correction is about $-0.2$\,dex at $\mathrm{[Fe/H]}<-3.0$.

We used 12 \ion{Sc}{2} lines in the wavelength range 3350\,-\,4420\,
{\AA} to determine the Sc abundance, finding $\mathrm{[Sc/Fe]}=+0.29$. We do not
take into account the effects of hyperfine splitting, which are expected to be
negligible in an analysis of weak lines. No NLTE calculations for Sc for
EMP stars are available in literature.

\subsection{Iron-Peak Elements: V, Cr, Mn, Co, Ni, Cu, and Zn}

Vanadium abundances have not been reported for many metal-poor stars,
because even the strongest \ion{V}{1} line in the optical at 4379\,{\AA}
is difficult to detect. The line is also not detected in our spectrum of
{\bd}, and we estimate the upper limit on the V abundance ratio to be
$\mathrm{[V/Fe]}<+0.20$. On the other hand, our near-UV spectrum permits
measurements from two detectable \ion{V}{2} lines at 3545\,{\AA} and
3592\,{\AA}. The derived V abundance ratio is $\mathrm{[V/Fe]}=-0.02$, which
is consistent with the result from the \ion{V}{1} line. We may need to
exercise caution when comparing our V abundance determined from the \ion{V}{2} with
those determined from \ion{V}{1} for other stars, given the result of
\cite{Johnson2002} that \ion{V}{2} lines yielded about 0.2\,dex higher
V abundances than that from \ion{V}{1} lines, although \cite{Lai08} did
not report such an offset.

The Cr abundance is measured from six \ion{Cr}{1} lines and two
\ion{Cr}{2} lines. The resultant abundance from \ion{Cr}{2} is
0.15\,dex higher than that obtained from \ion{Cr}{1}. Although measurements of
\ion{Cr}{2} lines are limited in previous studies of
metal-poor stars, as the detectable lines are located at the near-UV region
($<3500$\,{\AA}), several studies found similar discrepancies
\citep{Frebel07b,Lai08,Bonifacio09}. \cite{Johnson2002} reported a
similar offset associated with optical \ion{Cr}{2} lines for mildly
metal-poor stars. We refer to the \ion{Cr}{1} result as our final Cr
abundance ratio, $\mathrm{[Cr/Fe]}=-0.37$.

%These disagreements between neutral and ionized species may come from the NLTE effects. However, no such computation for V nor Cr exists.

We measure the Mn abundances from two \ion{Mn}{1} lines at 4030\,{\AA}
and three \ion{Mn}{2} lines in the wavelength range 3440\,-\,3490\,
{\AA}. The effects of hyperfine splitting are neglected because these lines
are quite weak. The \ion{Mn}{2} results are 0.3~dex higher than the \ion{Mn}{1}
results (Table~\ref{abundances}). \cite{Cayrel04} found that Mn
abundances derived from the \ion{Mn}{1} resonance triplet at 4033\,{\AA}
are systematically lower than those from other lines, and corrected
their Mn abundances from the triplet by +0.4\,dex. \cite{Frebel07b} and
\cite{Bonifacio09} also followed this procedure. \cite{Bergemann008}
suggested that the discrepancy is due to NLTE effects. Their NLTE
calculations indicate that the abundance corrections for the
\ion{Mn}{1} triplet are 0.2\,dex higher than those for other
\ion{Mn}{1} lines, although the difference does not reach
0.4\,dex. Thus, NLTE effects might resolve the 0.3\,dex disagreement
between the Mn abundance ratios determined for {\bd} from the \ion{Mn}{1}
and \ion{Mn}{2} lines. We here adopt the result from the \ion{Mn}{2}
lines, $\mathrm{[Mn/Fe]}=-0.79$.

A number of \ion{Co}{1} and \ion{Ni}{1} lines are used to obtain the
Co and Ni abundance ratios, respectively ($\mathrm{[Co/Fe]}=+0.54$,
$\mathrm{[Ni/Fe]}=+0.08$). The \ion{Co}{1} lines could be severely
affected by NLTE. According to \cite{Bergemann10}, the NLTE analysis
increases Co abundances by as much as 0.7\,dex at $\mathrm{[Fe/H]}=-3$
compared to the LTE analysis.

Thanks to our high-quality near-UV spectrum, we are able to measure two
\ion{Cu}{1} lines at 3248\,{\AA} and 3274\,{\AA}, resulting in an
abundance ratio of $\mathrm{[Cu/Fe]}=-0.91$.

Zn is not detected in the observed spectrum. From inspection of the
strongest \ion{Zn}{1} line at 3345\,{\AA}, we set an upper limit on its
abundance ratio of $\mathrm{[Zn/Fe]}<+0.17$.

\subsection{Neutron-Capture Elements: Sr, Y, Zr, Ba, Eu, and Pb}

The Sr abundance ratio is derived from the \ion{Sr}{2} resonance lines at
4077\,{\AA} and 4215\,{\AA}, $\mathrm{[Sr/Fe]}=-0.23$. NLTE
corrections for the \ion{Sr}{2} lines depend on the adopted stellar parameters,
and are positive for some stars, ane negative for others
\citep{Mashonkina001,Mashonkina08,Andrievsky11}. 
In any case, it appears that these corrections would not be large ($<0.1$\,dex).

%, although the NLTE calculations for
%metal-poor subgiants have not been conducted.

We determine the Y abundance ratio from the \ion{Y}{2} line at 3601\,{\AA},
and the Zr abundance ratio from the \ion{Zr}{2} lines at 3438\,{\AA} and
3552\,{\AA}. The results are $\mathrm{[Y/Fe]}=-0.24$, and
$\mathrm{[Zr/Fe]}=+0.04$. We clearly detect the \ion{Y}{2} line at
3774\,{\AA} but do not use it, because it overlaps with a nearby Balmer line.

The Ba abundance ratio, $\mathrm{[Ba/Fe]}=-0.60$, is derived from the
\ion{Ba}{2} lines at 4554\,{\AA} and 4934\,{\AA}. We take hyperfine
splitting into account, using the line list of \cite{McWilliam1998},
and assuming the isotope ratios of the $r$-process component of
Solar System material. \cite{Andrievsky09} performed NLTE calculations
for \ion{Ba}{2}, and obtained about 0.1\,dex higher Ba abundances than
LTE analyses for EMP dwarfs with
$T_\mathrm{eff}=5500\,\mathrm{K}$. For giants with the same
metallicity and temperature, the NLTE correction amounts to 0.2\,dex
or more. However, even if the NLTE effect is considered, Ba remains
under-abundant in {\bd}.

No Eu line is detected in the observed spectrum. From inspection of
the \ion{Eu}{2} line at 3820\,{\AA}, we set an upper limit of
$\mathrm{[Eu/Fe]}<+0.41$. The lower limit on the Ba/Eu ratio
([Ba/Eu]$>-1.0$) permits the origin of these heavy elements to be
associated with the $r$-process, as well as with the $s$-process.

\begin{figure}
\epsscale{.70}
%\plotone{Pb.eps}
\plotone{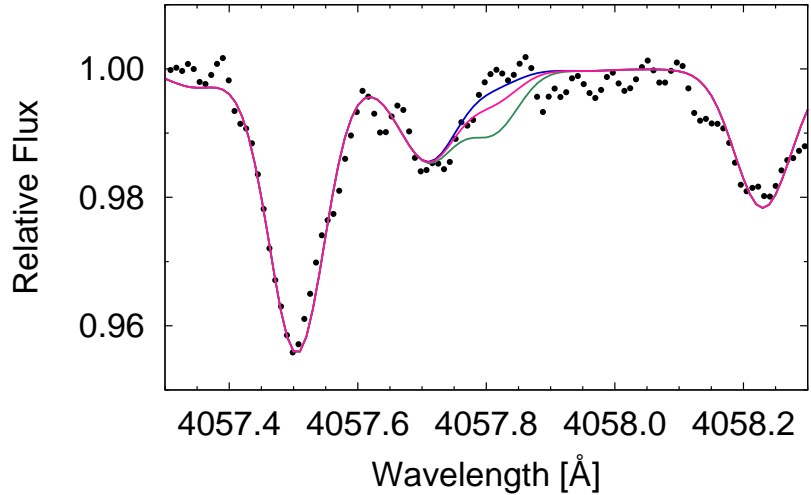}
\caption{Same as Figure~\ref{Lithium}, but for the \ion{Pb}{1} line at
4058\,{\AA}. Assumed abundances are $\log \epsilon(\mathrm{Pb})=-0.4$, $-0.1$, 
and $+0.2$. Note that, for values greater than $\log
\epsilon(\mathrm{Pb})=-0.1$, our adopted uppper limit, Pb should
clearly be detectable.}
\label{Pb}
\end{figure}

The upper limit on the Pb abundance, $\log
\epsilon(\mathrm{Pb})<-0.10$, is estimated from a comparison of the
observed spectrum with synthetic spectra in the region of the
4057.8\,{\AA} \ion{Pb}{1} line (Figure~\ref{Pb}). The line list
  for the Pb isotopes of \cite{vaneck03} is adopted, and Solar System Pb
  isotope ratios are assumed. In the spectrum synthesis, we modified
the C and Mg abundances, within their errors, so as to fit the CH 
and Mg lines neighboring the \ion{Pb}{1} line.

%\subsection{Comparison with Other Metal-Poor Stars}

\section{Discussion}\label{sec:disc}

\subsection{Origin of the Carbon Excess}

As shown by previous studies, CEMP-no stars are intriguing objects.
They tend to have lower metallicity than CEMP-{\it s} and most
carbon-normal stars, and are thus expected to offer vital clues for
understanding the chemical evolution of the early Universe. However, 
consensus on the nucleosynthetic origin of this class of stars has yet to
be reached. {\bd} is a particularly important star for addressing this
question, as it is by far the brightest CEMP-no star at extremely low
metallicity.  In this section, we examine the scenarios that have been
proposed to account for the origin of CEMP-no stars, by comparing the
obtained abundance pattern of {\bd} with theoretical predictions.

We first note that self-enrichment of carbon by dredge-up of the
products of the helium burning (e.g., \citealt{Fujimoto00}) is
clearly excluded, because {\bd} is an unevolved subgiant.

Mass transfer from a companion AGB star is proposed as one of the
causes of C enrichment for many CEMP stars, and such a model has had
great success in explaining the observed properties of CEMP-{\it s}
stars \citep[e.g.][]{kappeler11}. However, the observed properties of
{\bd} do not support this scenario. The first difficulty, which
applies to all CEMP-no stars (by definition), is that the
neutron-capture elements that are expected to be enhanced by an AGB
companion are not over-abundant in this star. \cite{Cohen06},
following \citet{Busso99}, suggested that the apparently normal Ba
abundances of CEMP-no stars might be explained by the high neutron-to-
Fe-peak-element seed ratio in the {\it s}-process that occurred in
the AGB companion, resulting in little or no Ba excess, but large Pb
enhancement.  This follows, according to \citet{Busso99}, because the
very large number of neutrons available per Fe-peak-element seed
nucleus at low metallicity forces the $s$-process to run to
completion, and terminate on lead. Under this hypothesis, the Pb
abundance predicted by \citet{Cohen06} is $\log \epsilon(\mathrm{Pb})
=1.5$ at [Fe/H] $=-3.5$, although they comment that its detection
would be difficult, because of the weakness of the line, and its
location in a spectral region contaminated by a ``thicket of CH
features."  In this regard, we are fortunate that {\bd} is as bright
as it is, since the high S/N spectrum we have obtained (as well as the
relatively high effective temperature) provides the opportunity to set
a meaningful upper limit, even from measurements in the optical.  Our
measured upper limit for lead in {\bd}, ($\log
\epsilon(\mathrm{Pb})<-0.10$), is clearly much lower than the
predicted value from \citet{Cohen06}, $\log \epsilon(\mathrm{Pb}) =1.5$.

\cite{Komiya07} discussed the possibility that, since a relatively high-mass 
companion AGB star ($M>3.5M_\odot$) produces little {\it s}-process
elements, due to inefficient radiative $^{13}$C burning, this might 
account for the lack of Ba enrichment in CEMP-no stars.  However, since a
high-mass AGB star converts C into N via hot-bottom burning, the low N
abundance of {\bd} permits only a narrow range of mass. Another
constraint is the low C/O ratio (C/O $<1$) found for {\bd}, which cannot
be explained by the AGB nucleosynthesis scenario \citep[e.g.,
][]{Nishimura09}. Moreover, the radial velocity of {\bd} has stayed
constant since 1984 (\S2.3), indicating no signature of
binarity (or only allowing a binary of very long period). Our conclusion
is that the binary mass-transfer scenario is excluded as the origin of
the C excess of {\bd}.

Another scenario that has been put forward, that the C enhancement occurs
prior to the formation of (at least some) CEMP stars due to the
predicted mass loss from rapidly-rotating massive stars of the lowest
metallicity. \cite{Meynet06} explored models of rapidly-rotating
$60M_\odot$ stars with total metallicities $Z=10^{-8}$ and $10^{-5}$, and
found that strong internal mixing increases the total metallicity at the
stellar surface significantly, leading to large mass loss in the form
of a vigorous wind. The ejecta are highly enriched in CNO elements,
which are products of triple-$\alpha$ reactions and the CNO cycle. In
particular, the N excess is predicted to be quite large, due to
operation of the CNO cycle in the H-burning shell, which converts C into
N. The low observed N abundance of {\bd}, however, cannot be
accounted for by this scenario. Furthermore, recent models of rotating
massive stars with $Z=0$ explored by \cite{Ekstrome08}, over a wide
range of mass (9\,-\,200$M_\odot$), suggest that these stars experience
only relatively little mass loss, in stark contrast to that from
extremely low, but non-zero, metallicity models. Their results indicate
that the very first generation of stars, even though they might be
rapidly rotating, are unlikely to lead to CNO enrichment through mass
loss.

A remaining possibility is element production by so-called faint
supernovae associated with the first generations of stars, which
experience extensive mixing of matter and fallback onto the nascent
compact object that forms at their centers during their
explosions \citep{Umeda03,Umeda05,Tominaga07b}. Such a process is
apparently realized in relativistic jet-induced supernovae with low
energy-deposition rates \citep{Tominaga07a}. Since a relatively large amount of
material collapses onto the central remnant in such supernovae, only
small amounts of heavy elements, such as the Fe-peak elements (which are
synthesized in the inner regions of the progenitors) can be ejected,
compared to lighter elements, such as C, which are synthesized in the 
outer regions. Thus, high [C/Fe] and [O/Fe] ratios are predicted in the
ejected material. Since the luminosity of supernovae is generated by the
radioactive decay of $^{56}$Ni through $^{56}$Co to $^{56}$Fe, the
small amount of $^{56}$Ni ejected  results in the faintness of the explosion.

\begin{figure}
%\plotone{apBD+44_493.eps}
\plotone{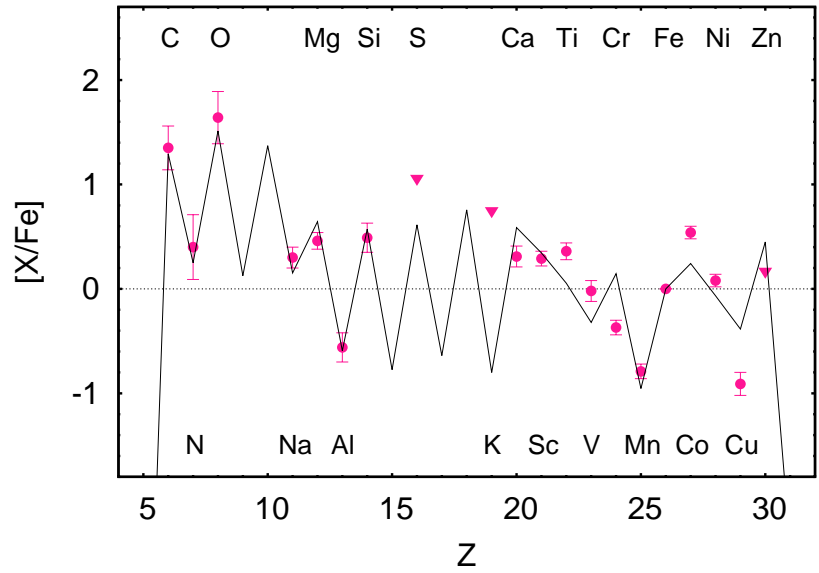}
\caption{Comparison between the observed elemental abundance pattern
  of {\bd} (filled circles) and the theoretical individual supernova
  yields (solid lines). The upper limits on the abundances for several elements
    by the observation are shown by triangles. See text for details
  of the supernova model.}
\label{ap-SN}
\end{figure}

This scenario does not produce a serious conflict with the observed 
elemental-abundance pattern of {\bd}. Indeed, a faint Pop
III supernova model which undergo mixing-and-fallback during the 
explosion can well-reproduce the abundance pattern of this 
object, as shown in Figure~\ref{ap-SN}. 
This model is constructed with the same method as in \citet{Tominaga07b}
for a main-sequence mass of 25Msun \citep{Umeda05}, and explosion energy of $5\times10^{51}$\,ergs, while a normal 
supernova explodes with an energy of $\sim 1\times10^{51}$\,ergs; the ejected $^{56}$Ni mass is only about 10\% of that for a normal supernova. 
Since a supernova
with high entropy leads to a large [Zn/Fe] ratio \citep{Umeda05,Tominaga07b},
our low upper limit for [Zn/Fe] for {\bd} may constrain the explosion
energy, and favor a supernova with normal entropy. The low [N/C] in {\bd}
indicates that mixing between the He convective shell and H-rich
envelope during pre-supernova evolution is not significant
\citep{Iwamoto05}, and in the calculation only 15\% of the C in the He layer
is assumed to be converted into N, in order to reproduce the abundance
pattern. Thus, at present, a faint supernova is the most promising
candidate for the origin of the C excess in {\bd}.

%This scenario does not produce a serious conflict with the observed
%elemental-abundance pattern of {\bd}. Indeed, a faint Pop III supernova
%model can well-reproduce the abundance pattern of this object, as shown
%in Figure~\ref{ap-SN}. The nucleosynthesis yield for this model
%is calculated for a main-sequence mass of $25M_\odot$, and explosion
%energy of $5\times10^{51}$\,ergs, while a normal supernova explodes with
%an energy of $\sim 1\times10^{51}$\,ergs; the ejected $^{56}$Ni mass is
%only about 10\% of that for a normal supernova. 

Our conclusion on the origin of the C excess in {\bd} impacts
the interpretation of the entire CEMP-no class of stars. Faint
supernovae should have played an important role in the chemical
evolution of the early Galaxy, and at least some (perhaps all) CEMP-no
stars probably formed from gas polluted by them. This supports the
notion that the ultra metal-poor (UMP; [Fe/H] $< -4.0$) and hyper
metal-poor stars (HMP; [Fe/H] $< -5.0$) are not low-mass Pop III stars
but, rather, Pop II stars that formed from gas that had been
pre-enriched by high-mass Pop III stars.

\citet{norris12} discussed the possible existence of two cooling
channels responsible for the formation of second-generation stars from
gas clouds polluted by first generations of massive stars, following the
studies by \citet{bromm03} and \citet{frebel07a}. One of them is the
process responsible for non carbon-enhanced stars, which is not yet
well-defined; dust-induced cooling is a candidate. The other is the
cooling by fine-structure lines of \ion{C}{2} and \ion{O}{1} in the case
of carbon and oxygen-enhanced material. Our conclusion that {\bd} is
formed from a gas cloud pre-enriched in carbon supports the existence of
this channel of EMP star formation.

\subsection{Production and Depletion of Light Elements}

\begin{figure}
\epsscale{.50}
%\plotone{Be-Fe.eps}
%\plotone{Be-O.eps}
\plotone{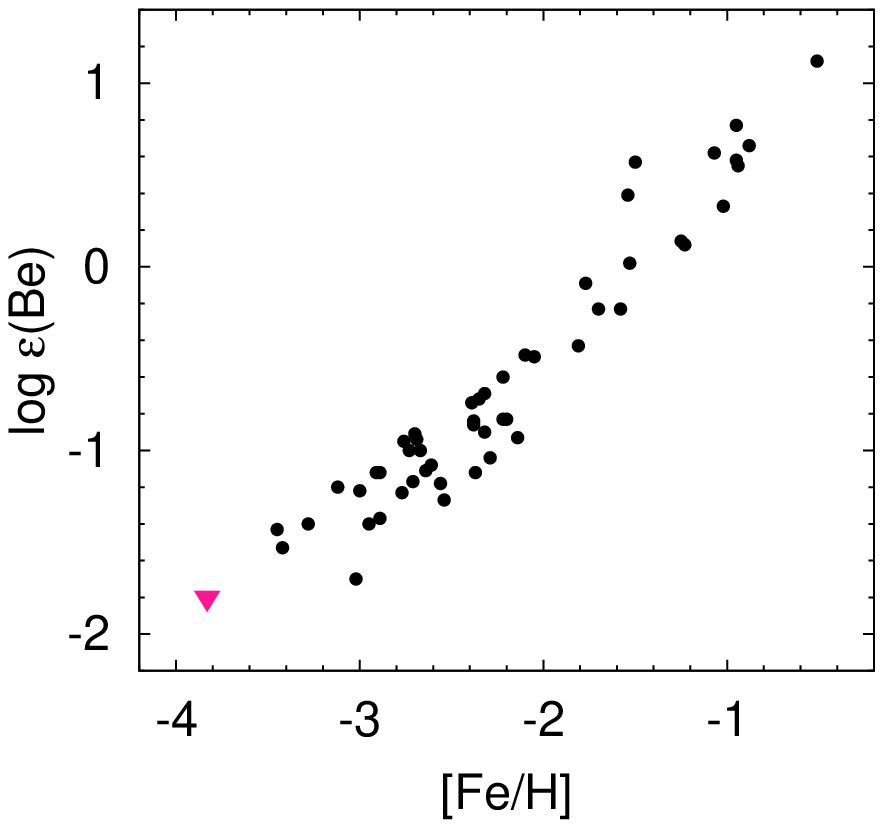}
\plotone{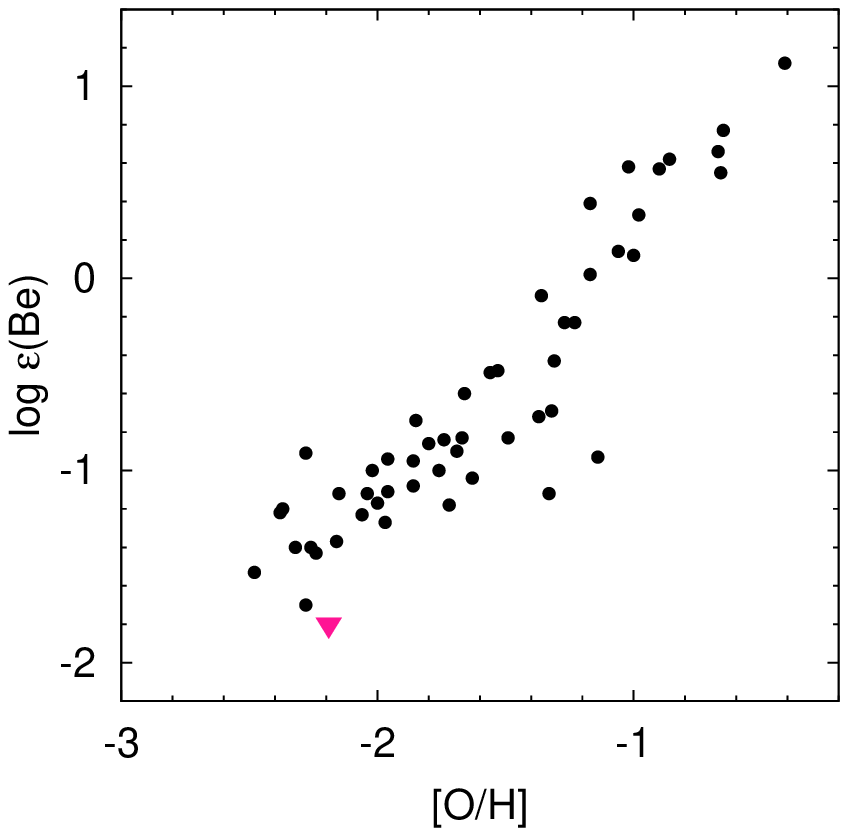}
\caption{Be abundances, as a function of [Fe/H] (upper panel) and [O/H] (lower panel). The upper limit for {\bd} is shown by the filled triangle. Filled circles represent the result of \cite{Rich009}.}
\label{Be-Fe}
\end{figure}

Figure~\ref{Be-Fe} shows the Be abundance upper limit for {\bd} in the
$\log \epsilon(\mathrm{Be})$ vs. [Fe/H] plane, along with the results
from \cite{Rich009}. Our observation (an upper limit) is consistent
with the linear trend between Be and Fe seen in this figure. It is of
significance that the Be abundance keeps decreasing and exhibits no
plateau at low metallicity, at least to the level of $\log
\epsilon(\mathrm{Be})<-1.8$. Previous measurements of Be for stars
permitted the possibility of a Be plateau around $\log
\epsilon(\mathrm{Be}) \sim-1.4$ \citep{Primas00a, Primas00b}, which
would now be clearly called into question.

Our analysis is the first attempt to measure a Be abundance for a CEMP
star. Since Be is produced via the spallation of CNO nuclei, CNO
abundances, especially O abundances, have been expected to correlate
with Be abundances. However, our low Be upper limit shows that the high
C and O abundances in {\bd} are irrelevant to its Be abundance
(Figure~\ref{Be-Fe}, lower panel). Moreover, the origin of the C excess in
CEMP-no stars such as {\bd}, which we have argued is most likely due to
faint supernovae, is unlikely to be a significant source of high-energy
CNO nuclei that participate in the primary spallation process expected
to be associated with Be production.

Be abundances, as well the abundances of B, in metal-poor stars are proposed to
be good indicators of the ages of stars formed in the early Galaxy
\citep{suzuki99, beers00, smiljanic10}, because they could reflect the
increase of cosmic ray flux in the early Galaxy, while other metals such
as Fe would reflect the contributions from individual supernovae, and
result in a spatial inhomogeneity in the abundance ratios. Our finding
that Be is not significantly enhanced even by the supernovae that yield
large amounts of C and O, provides additional support for this
hypothesis.

%SOME WORDS ON THE LOW BE ABUNDANCE AND IMPLICATIONS FOR THE AGE OF THIS
%STAR, SINCE IT MUST HAVE BEEN FORMED AT A TIME WHEN THE COSMIC RAY FLUX
%IN THE EARLY GALAXY WAS VERY LOW (REFER TO BEERS ET AL. 2000 IAU SYMP
%98 AND SMILJANIC ET AL. (2010 IAU SYMP 268 AND REFS THEREIN).
%\subsection{Lithium and Beryllium Depletion}

The above discussion is, however, based on the assumption that the Be
abundance in {\bd} is not depleted during its evolution. Beryllium is a
fragile element, with a burning temperature of $\sim 3.0\times10^6\,
\mathrm{K}$, higher than that of Li ($2.5\times10^6\,
\mathrm{K}$). Given that the Li abundance of this star, 
$\log \epsilon(\mathrm{Li})=1.0$, is much lower than the Spite plateau value,
we should also examine the possibility of Be depletion in {\bd}.

\begin{figure}
\epsscale{.70}
%\plotone{Li-Teff.eps}
\plotone{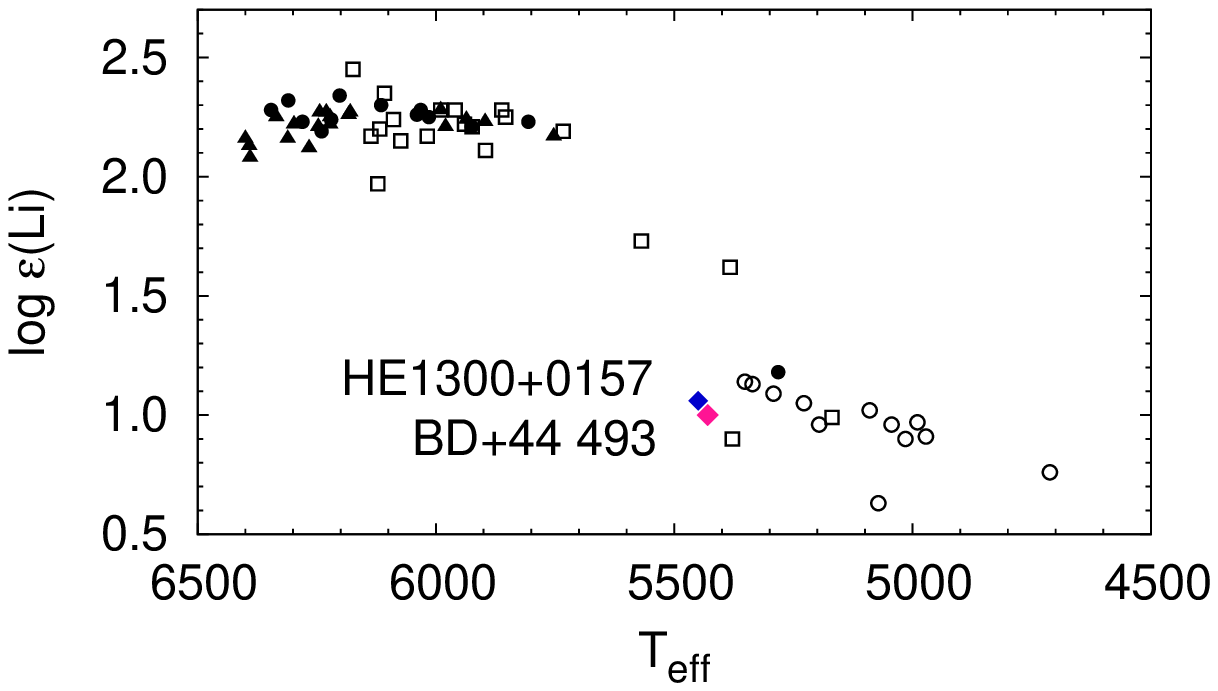}
\caption{Li abundances for metal-poor ($\mathrm{[Fe/H]}<-1.4$) dwarfs
  and subgiants, as a function of effective temperature. Our result for
  {\bd} is shown by the (pink) diamond, whereas the (blue) diamond
  indicates HE~1300+0157 \citep{Frebel07b}. Open circles represent the
  results of \cite{GarciaPerez06}. Open squares, filled circles, and
  filled triangles were taken from \cite{Boesgaard06}, \cite{Shi07},
  and \cite{Asplund06}, respectively. For
  $T_\mathrm{eff}<5700\,\mathrm{K}$, only subgiants ($\log g<4.0$) are
  plotted.}
\label{Li-Teff}
\end{figure}

In Figure~\ref{Li-Teff}, we compare the Li abundance of {\bd} with those
of other metal-poor ($\mathrm{[Fe/H]}<-1.4$) subgiants, which are mainly
taken from \cite{GarciaPerez06}, as well as metal-poor dwarfs with
$T_\mathrm{eff}>5700\,\mathrm{K}$. As a star evolves from the main
sequence to the red giant branch, the effective temperature decreases,
that is, cooler subgiants in this diagram are more evolved. At
$T_\mathrm{eff}>5700\,\mathrm{K}$, metal-poor subgiants appear to have
Li abundances consistent with the Spite plateau values, along with the
metal-poor dwarfs. The sharp decline of Li abundances in the range
5700\,-\,5400\,K indicates that significant Li depletion occurs when
stars pass through this temperature range, often referred to as the
Boesgaard dip \citep{Boesgaard02}. \cite{GarciaPerez06} concluded
that their subgiants have depleted Li abundances that are generally
consistent with predictions from standard models of stellar evolution
\citep{Deliyannis90}. Since the Li abundance of {\bd} is in agreement
with the values found for their sample, we conclude that the Li
depletion in {\bd} is likely to be similar to those found in their
sample, even though the metallicity is remarkably lower.

\begin{figure}
\epsscale{.70}
%\plotone{Be-subgiants.eps}
\plotone{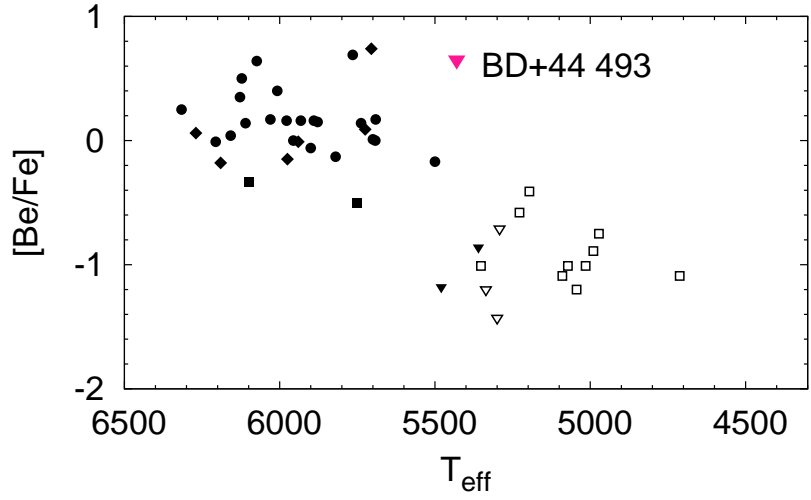}
\caption{[Be/Fe] ratios,  as a function of effective temperature for metal-poor
  ($\mathrm{[Fe/H]}<-1.4$) subgiants ($\log g<4.0$). Our upper limit
  for {\bd} is shown by the pink triangle. Open symbols indicate the
  results of \cite{GarciaPerez06}. Filled symbols were taken from
  \cite{Rich009}, \cite{Smiljanic09}, and \cite{Tan09}. All triangles
  mean upper limits.}
\label{Be-subgiants}
\end{figure}

In order to investigate Be depletion, we adopt [Be/Fe] as an indicator.
We assume that metal-poor stars originally have Be abundances in
proportion to their Fe abundances, as found in Figure~\ref{Be-Fe}, that
is, [Be/Fe] is close to 0.2. If Be is not depleted in a subgiant, the
star is expected to follow the linear correlation, and have the same
[Be/Fe] value as dwarfs, while lower [Be/Fe] values suggest Be depletion
in the star. In Figure~\ref{Be-subgiants}, we plot [Be/Fe] versus
effective temperature for metal-poor subgiants ($\mathrm{[Fe/H]}<-1.4$;
the same metallicity range as adopted for discussion on Li depletion).
It appears that objects with $T_\mathrm{eff}>5700\,\mathrm{K}$ did not
experience Be depletion, as expected from our discussion on the Li
abundances in the temperature range. On the other hand, [Be/Fe] values
in cool subgiants are about one order of magnitude lower, which could
be interpreted as the result of Be depletion.

During the subgiant phase, the stellar convection zone extends into the
interior, first reaching to the Li-depleted layer, and later, to the 
Be-depleted layer, because nuclear Be burning requires higher
temperatures than Li burning. Therefore, Be depletion is expected in
more-evolved subgiants with lower effective temperatures than Li depletion.
However, this is not clearly found in Figures~\ref{Li-Teff} and
\ref{Be-subgiants}. One reason might be the paucity of stars of the sample in
the temperature range 5400\,-\,5700\,K. More measurements of Li and
Be abundances for metal-poor subgiants in this temperature range are
strongly desired in order to assess this issue. Futhermore, the
uncertainty in effective temperature determinations could make the
difference unclear. In comparisons with the results of previous studies, 
consistency of temperature determinations are essential. Since {\bd}
has an effective temperature of 5430\,K that falls into the range where
Be depletion appears to occur, it is difficult to conclude whether or
not Be is depleted in this star.

Another constraint on Be depletion could be obtained from boron
abundance measurements. If B is depleted in a star, Be also must be
depleted, because the burning temperature of B,
$5.0\times10^6\,\mathrm{K}$, is higher than that of Be. The absence of
an available B line in the wavelength range that can be accessed from
ground-based telescopes makes a B abundance measurement a
challenge. The STIS (Space Telescope Imaging Spectrograph) instrument
onboard the Hubble Space Telescope is at present the only spectrograph with
which we could observe the \ion{B}{2} resonance lines at
2497\,{\AA}. Abundance measurements of Li, Be, and B for
metal-poor subgiants will offer important clues for our understanding
of mixing processes in stellar interiors, which can decrease surface
abundances of light elements. Theoretical modeling of the depletion
for a wide range of metallicity and stellar mass should also be
explored.

\section{Summary and Concluding Remarks}

We have conducted a detailed elemental abundance analysis for the 9th
magnitude EMP star {\bd}, based on very high-quality, high-resolution
optical and near-UV spectra. This object is a CEMP-no star, and has
exhibited no detectable radial-velocity variations over the past 24 years.
The abundance patterns we measure, in particular C, N,
O, as well as the heavy neutron-capture elements, suggest that the most
likely origin of this object is that it was born from gas polluted by
the elements produced by a faint supernova undergoing mixing and
fallback, which ejected only a small amount of metals beyond C, N,
and O.

Recent searches for EMP stars have revealed that the fraction of CEMP
stars among them is substantially higher than that for less metal-poor
stars \citep{Carollo12}, and that CEMP-no stars are an important
contributor to these \citep[e.g., ][]{Aoki12,norris12}. The HMP stars
HE~1327--2326 and HE~0107--5240 are included in this class of objects.
The results for {\bd} suggest a very important role for faint supernovae
in the earliest phase of chemical enrichment in the Galaxy, and (at
least for some low-mass stars) the importance of cooling by
fine-structure lines of \ion{C}{2} and \ion{O}{1} in their early formation.

The low observed Li abundance and non-detection of Be for {\bd} also shed new
light on the production of light elements in the early Galaxy.  However,
it remains possible that these elements could have been 
depleted in this star during the course of its evolution. A
future (space-based) measurement of B for this star has particular importance to
distinguish whether these effects apply or not. Further measurements of Li and Be for
subgiants with {\Teff} of about 5500~K, in which depletion of these
elements first appears, are also useful for understanding the nature
of light elements in EMP stars.

Finally, we emphasize that our measured upper limit on the abundance of
Pb ($\log \epsilon$(Pb)$< -0.10$) rejects that possibility that the
observed neutron-capture-element abundance patterns in {\bd} arise from
the operation of a modified $s$-process at low metallicity in CEMP-no
stars. High-resolution spectroscopy in the region of the UV Pb line of
{\bd} from HST/STIS has recently been obtained by Beers et al. (in
preparation), and can be used to confirm our estimated limit. Taken at
face value, our present result already provides additional strong
evidence that the C and O enhancements in {\bd} must have originated in
an astrophysical site other than AGB stars, as discussed in \S5.1.
Additional measurements of Pb from the ground and with HST/STIS would be
highly desirable to obtain for low-metallicity CEMP-no stars, if other
sufficiently bright examples are identified in the future.
 
%, as had been suggested previously by \citep{Cohen06}, who
%predicted an abundance of log$\epsilon$(Pb) $\sim +1.5$ for CEMP-no
%stars with [Fe/H] $= -3.5$. It is fortunate that {\bd} is sufficiently
%bright for a measurement of the Pb limit to be obtained from the optical
%Pb line, despite the influence of nearby CH lines. 

W.A. and N.T. were supported by the JSPS Grants-in-Aid for Scientific
Research (23224004).  T.C.B. acknowledges partial funding of
this work from grants PHY 02-16783 and PHY 08-22648: Physics Frontier
Center/Joint Institute for Nuclear Astrophysics (JINA), awarded by the
U.S. National Science Foundation.

{\it Facilities:} \facility{Subaru (HDS)}

\appendix
\section{A spectral atlas of {\bd}}

{\bd} is, by far, the brightest object among EMP stars found to date.
As such, the very high-quality spectrum of this star analyzed in the
present work can serve as a useful template for future studies of EMP
and CEMP stars. Since this object is a CEMP-no star, it displays clear
excesses of C and O, and is at an effective temperature ({\Teff} =
5430~K) such that many CH and OH features are identifiable in the
near-UV and blue spectral regions.

\begin{figure}
\epsscale{.90}
%\plotone{bd44ex.eps}
\plotone{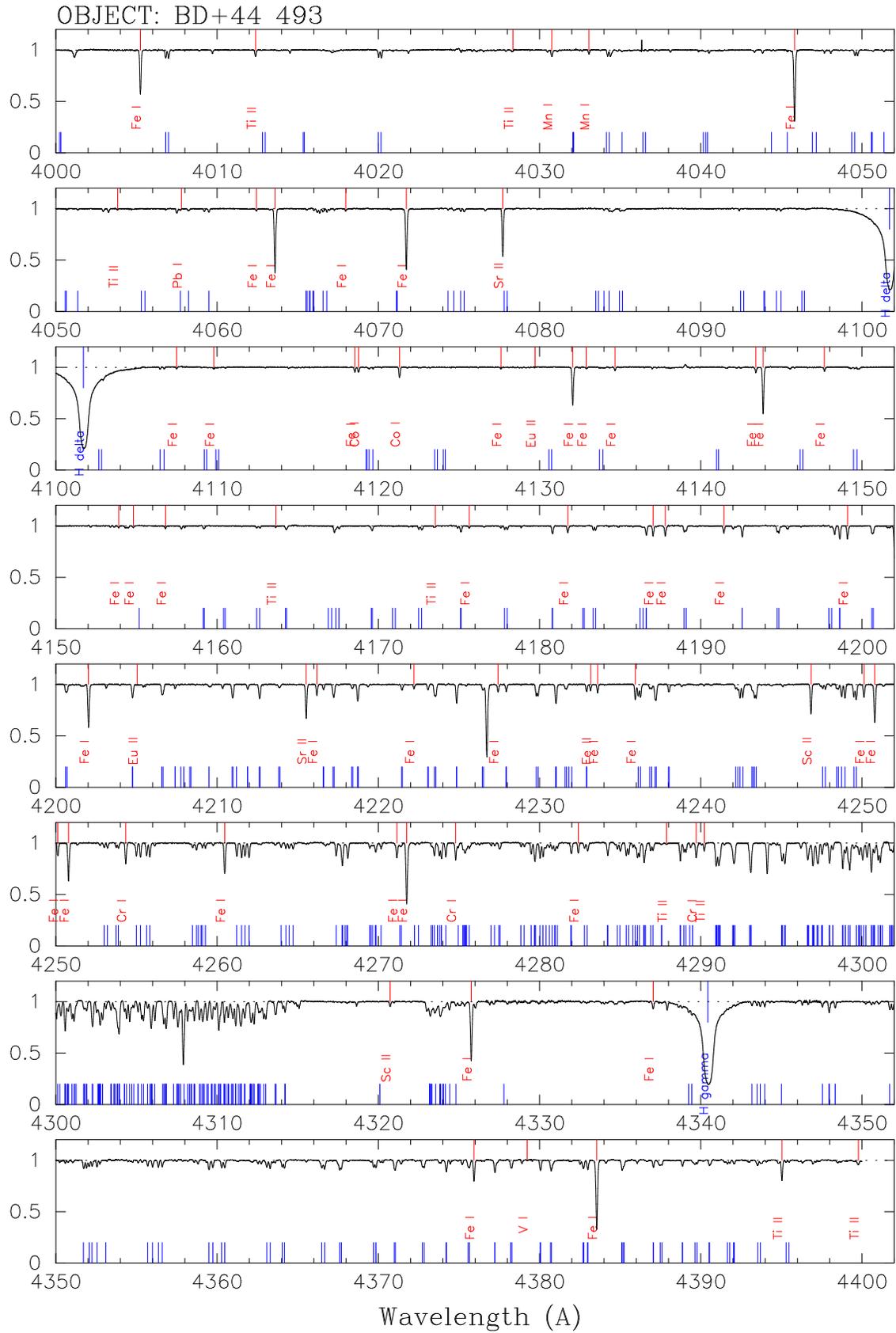}
\caption{A portion of the spectral atlas of {\bd}.  The full set of
plots are provided online only.}
\label{fig:atlas}
\end{figure}

We provide an atlas of the spectrum of {\bd} as online-only figures.
Figure~\ref{fig:atlas} shows a portion of the atlas. The full spectrum
covers the wavelength range from 3080~{\AA} to 9370~{\AA}. A small
portion of this wavelength region is not covered, due to the gap in the
two CCDs in HDS (5330--5430~{\AA}), and in the redder regions where the
limited size of the CCDs does not fully cover the free spectral range
($>7000$~{\AA}), as well as due to the occasional bad columns of the CCDs. Some spectral
regions are affected by telluric absorption that is not fully corrected.
The wavelength range between 3900--3960~{\AA} is not covered by the
spectrum obtained with resolving power of $R=90,000$, but it is
supplemented by the $R=60,000$ spectrum (see \S~\ref{sec:obs}).

The atomic lines used in the abundance analysis in the present work
are indicated by the shorter (red) lines extended down from the top of each panel.
Other atomic lines that were not used in the analysis, including hydrogen lines,
are shown by the longer (blue) lines. Molecular features used in this
analysis are shown by lines extending up from the bottom of each panel
for OH (long, blue), CH (short, red) and NH (middle, green).

\clearpage

%\begin{figure}
%\epsscale{.70}
%\caption{Same as Figure~\ref{Lithium}, but for CH features at 4324\,{\AA}. Assumed abundances are $\mathrm{[C/Fe]} =+1.10$, +1.30, and +1.50.}
%\label{CH4324}
%\end{figure}

%\begin{figure}
%\epsscale{.70}
%
%\caption{Same as Figure~\ref{Lithium}, but for CH features at 3144\,{\AA}. Assumed abundances are $\mathrm{[C/Fe]} =+1.05$, +1.25, and +1.45.}
%\label{CHUV}
%\end{figure}

\clearpage

\begin{deluxetable}{lllccr}
\tablewidth{0pc}
\tablecaption{Subaru/HDS Observations of BD+44$^\circ$493 and HD~12534 \label{obslog}}
\tablehead{
\colhead{Target} & \colhead{Universal Time\tablenotemark{a}} &
\colhead{Wavelength\tablenotemark{b}} & \colhead{Resolving} & \colhead{CCD Binning} & \colhead{Total Time} \\
& & \colhead{({\AA})} &\colhead{Power} & \colhead{(pixels)} & \colhead{(sec)}
}
\startdata
{\bd} & 2008 Aug 22, 15:19 & 4030--5330,\,5430--6730 & 60,000 & $2\times2$ & 600 \\
{\bd} & 2008 Oct 4, 09:00 & 4030--5330,\,5430--6770 & 90,000 & $1\times1$ & 1200 \\
{\bd} & 2008 Oct 4, 09:28 & 6690--7960,\,8080--9370\tablenotemark{c} & 90,000 & $1\times1$ & 1500 \\
{\bd} & 2008 Oct 5, 10:06 & 3080--3900,\,3960--4780 & 90,000 & $1\times1$ & 7200 \\
{\bd} & 2008 Nov 16, 05:37 & 3540--4350,\,4420--5240 & 60,000 & $2\times2$ & 300 \\
HD12534 & 2008 Oct 4, 08:53 & 4030--5330,\,5430--6770 & 90,000 & $1\times1$ & 90 \\
HD12534 & 2008 Oct 4, 09:57 & 6690--7960,\,8080--9370 & 90,000 & $1\times1$ & 150 \\
HD12534 & 2008 Oct 5, 09:58 & 3080--3900,\,3960--4780 & 90,000 & $1\times1$ & 135 \\
\enddata
\tablenotetext{a}{UT shown is at the beginning of the observations.}
\tablenotetext{b}{The two wavelength regions correspond to the blue and
red CCDs, respectively.}
\tablenotetext{c}{The wavelength region is not fully covered, due to the limited CCD size.}
\end{deluxetable}

%\input{linelist}
% \begin{deluxetable}{lrrrr}
%\tablewidth{0pc}
%\tablecaption{Line Data and Equivalent Widths\label{tab:lines}}
%\tablehead{
% Species & Wavelength & $\log gf$ & l.e.p. & $W$ \\
%         & {\AA}        &         & (eV)    & m{\AA}   
%}
%\startdata
%  Na I  &   5889.951 &      0.117 &    0.000 &      66.03\\
%  Na I  &   5895.924 &     -0.184 &    0.000 &      45.26\\
%  Mg I  &   3829.355 &     -0.208 &    2.707 &      92.04\\
%  Mg I  &   3832.304 &      0.270 &    2.710 &     113.44\\
%  Mg I  &   3838.292 &      0.490 &    2.715 &     125.23\\
% \nodata & \nodata& \nodata& \nodata& \nodata \\
%\enddata
%\end{deluxetable}
 \begin{deluxetable}{lrrrrl}
\tablewidth{0pc}
\tablecaption{Line Data and Equivalent Widths\label{tab:lines}}
\tablehead{
 Species & Wavelength & $\log gf$ & l.e.p. & $W$ & Remarks\tablenotemark{a} \\
         & {\AA}        &         & (eV)    & m{\AA}   &
}
\startdata
  Na I  &   5889.951 &      0.117 &    0.000 &      66.03& \\
  Na I  &   5895.924 &     -0.184 &    0.000 &      45.26& \\
  Mg I  &   3829.355 &     -0.208 &    2.707 &      92.04& \\
  Mg I  &   3832.304 &      0.270 &    2.710 &     113.44& \\
  Mg I  &   3838.292 &      0.490 &    2.715 &     125.23& \\
  Mg I  &   4571.096 &     -5.393 &    0.000 &       3.17& \\
  Mg I  &   4702.991 &     -0.380 &    4.330 &      12.65& \\
  Mg I  &   5172.684 &     -0.450 &    2.712 &      97.88& \\
  Mg I  &   5183.604 &     -0.239 &    2.717 &     111.01& \\
  Mg I  &   5528.405 &     -0.341 &    4.346 &      11.56& \\
  Al I  &   3961.529 &     -0.336 &    0.014 &      42.38& \\
  Si I  &   3905.522 &     -1.090 &    1.909 &      82.35& \\
  Si I  &   3905.522 &     -1.090 &    1.909 &      82.35& \\
  Si I  &   3905.522 &     -1.090 &    1.909 &      82.35& \\
   V II &   3545.190 &     -0.390 &    1.100 &       1.76& \\
   V II &   3592.030 &     -0.370 &    1.100 &       1.69& \\
  Ti I  &   3653.490 &      0.220 &    0.050 &       7.58& \\
  Ti I  &   3989.759 &     -0.140 &    0.021 &       3.49& \\
  Ti I  &   3998.636 &      0.000 &    0.048 &       5.24& \\
  Ti I  &   4533.249 &      0.530 &    0.848 &       2.97& \\
  Ti I  &   4534.776 &      0.340 &    0.836 &       1.90& \\
  Ti I  &   4981.731 &      0.560 &    0.848 &       3.31& \\
  Ti I  &   5007.210 &      0.168 &    0.818 &       2.28& \\
  Ti II &   3148.050 &     -1.200 &    0.000 &      37.23& \\
  Ti II &   3190.880 &      0.190 &    1.080 &      38.18& \\
  Ti II &   3222.840 &     -0.480 &    0.010 &      66.56& \\
  Ti II &   3229.200 &     -0.550 &    0.000 &      68.46& \\
  Ti II &   3236.580 &      0.230 &    0.030 &      82.99& \\
  Ti II &   3239.040 &      0.060 &    0.010 &      78.88& \\
  Ti II &   3241.990 &     -0.050 &    0.000 &      75.44& \\
  Ti II &   3261.620 &      0.080 &    1.230 &      33.28& \\
  Ti II &   3278.290 &     -0.210 &    1.230 &      15.53& \\
  Ti II &   3282.320 &     -0.290 &    1.220 &      12.46& \\
  Ti II &   3287.660 &      0.340 &    1.890 &      15.09& \\
  Ti II &   3302.110 &     -2.330 &    0.150 &       4.89& \\
  Ti II &   3321.700 &     -0.320 &    1.230 &      14.52& \\
  Ti II &   3322.940 &     -0.090 &    0.150 &      70.79& \\
  Ti II &   3326.780 &     -1.180 &    0.110 &      34.93& \\
  Ti II &   3329.450 &     -0.270 &    0.140 &      65.21& \\
  Ti II &   3332.110 &     -0.150 &    1.240 &      45.49& \\
  Ti II &   3335.200 &     -0.440 &    0.120 &      60.23& \\
  Ti II &   3340.360 &     -0.610 &    0.110 &      57.25& \\
  Ti II &   3343.760 &     -1.270 &    0.150 &      31.66& \\
  Ti II &   3349.040 &      0.470 &    0.610 &      68.20& \\
  Ti II &   3361.220 &      0.280 &    0.030 &      94.82& \\
  Ti II &   3372.800 &      0.270 &    0.010 &      88.52& \\
  Ti II &   3380.280 &     -0.570 &    0.050 &      63.88& \\
  Ti II &   3383.770 &      0.140 &    0.000 &      86.13& \\
  Ti II &   3387.850 &     -0.430 &    0.030 &      66.91& \\
  Ti II &   3394.580 &     -0.540 &    0.010 &      63.60& \\
  Ti II &   3456.390 &     -0.230 &    2.060 &       3.15& \\
  Ti II &   3477.190 &     -0.970 &    0.120 &      44.47& \\
  Ti II &   3489.740 &     -1.920 &    0.140 &       8.43& \\
  Ti II &   3491.070 &     -1.060 &    0.110 &      39.03& \\
  Ti II &   3573.730 &     -1.500 &    0.570 &       7.35& \\
  Ti II &   3641.330 &     -0.710 &    1.240 &      11.77& \\
  Ti II &   3685.190 &     -0.040 &    0.570 &      72.47& \\
  Ti II &   3814.580 &     -1.700 &    0.570 &      11.79& \\
  Ti II &   3900.551 &     -0.280 &    1.130 &      24.96& \\
  Ti II &   3913.468 &     -0.410 &    1.120 &      21.83& \\
  Ti II &   4012.390 &     -1.610 &    0.570 &       6.19& \\
  Ti II &   4028.343 &     -0.990 &    1.892 &       2.34& \\
  Ti II &   4053.834 &     -1.060 &    1.893 &       1.18& \\
  Ti II &   4163.648 &     -0.210 &    2.590 &       1.43& \\
  Ti II &   4173.537 &     -1.815 &    1.084 &       2.43& \\
  Ti II &   4287.872 &     -1.820 &    1.080 &       1.64& \\
  Ti II &   4290.219 &     -0.930 &    1.165 &       7.85& \\
  Ti II &   4395.033 &     -0.460 &    1.084 &      20.93& \\
  Ti II &   4399.772 &     -1.220 &    1.237 &       7.42& \\
  Ti II &   4417.719 &     -1.190 &    1.165 &       4.17& \\
  Ti II &   4443.794 &     -0.720 &    1.080 &      15.75& \\
  Ti II &   4464.450 &     -1.810 &    1.161 &       1.58& \\
  Ti II &   4468.507 &     -0.600 &    1.131 &      15.40& \\
  Ti II &   4501.273 &     -0.770 &    1.116 &      13.24& \\
  Ti II &   4533.969 &     -0.540 &    1.237 &      13.03& \\
  Ti II &   4563.761 &     -0.690 &    1.221 &       9.44& \\
  Ti II &   4571.968 &     -0.320 &    1.572 &      10.62& \\
  Ti II &   4589.958 &     -1.620 &    1.237 &       1.49& \\
  Sc II &   3353.720 &      0.250 &    0.310 &      20.00& \\
  Sc II &   3535.710 &     -0.470 &    0.310 &       5.89& \\
  Sc II &   3572.526 &      0.267 &    0.022 &      46.07& \\
  Sc II &   3576.340 &      0.007 &    0.008 &      26.50& \\
  Sc II &   3580.925 &     -0.149 &    0.000 &      21.27& \\
  Sc II &   3590.470 &     -0.550 &    0.020 &       9.56& \\
  Sc II &   3630.740 &      0.220 &    0.010 &      35.69& \\
  Sc II &   3645.310 &     -0.420 &    0.020 &      13.53& \\
  Sc II &   4246.837 &      0.240 &    0.315 &      29.66& \\
  Sc II &   4320.732 &     -0.252 &    0.605 &       5.00& \\
  Sc II &   4400.399 &     -0.540 &    0.605 &       3.39& \\
  Sc II &   4415.563 &     -0.670 &    0.595 &       2.13& \\
  Cr I  &   3593.480 &      0.310 &    0.000 &      29.60& \\
  Cr I  &   4254.332 &     -0.114 &    0.000 &      21.27& \\
  Cr I  &   4274.796 &     -0.231 &    0.000 &      19.32& \\
  Cr I  &   4289.716 &     -0.361 &    0.000 &      16.35& \\
  Cr I  &   5206.038 &      0.019 &    0.941 &       5.97& \\
  Cr I  &   5208.419 &      0.158 &    0.941 &       7.49& \\
  Cr II &   3118.650 &     -0.100 &    2.420 &      23.35& \\
  Cr II &   3408.760 &     -0.040 &    2.480 &      14.41& \\
  Mn I  &   4030.763 &     -0.470 &    0.000 &       7.11& \\
  Mn I  &   4033.060 &     -0.618 &    0.000 &       4.30& \\
  Mn II &   3441.990 &     -0.273 &    1.780 &      11.45& \\
  Mn II &   3460.320 &     -0.540 &    1.810 &       5.93& \\
  Mn II &   3488.680 &     -0.860 &    1.850 &       2.25& \\
  Fe I  &   3225.790 &      0.380 &    2.400 &      27.36& \\
  Fe I  &   3286.750 &     -0.170 &    2.180 &      21.74& \\
  Fe I  &   3407.460 &     -0.020 &    2.180 &      26.08& \\
  Fe I  &   3440.610 &     -0.670 &    0.000 &      92.40& \\
  Fe I  &   3440.990 &     -0.960 &    0.050 &      82.79& \\
  Fe I  &   3443.880 &     -1.370 &    0.090 &      69.59& \\
  Fe I  &   3475.450 &     -1.050 &    0.090 &      79.44& \\
  Fe I  &   3476.700 &     -1.510 &    0.120 &      63.99& \\
  Fe I  &   3490.570 &     -1.110 &    0.050 &      78.85& \\
  Fe I  &   3497.840 &     -1.550 &    0.110 &      63.13& \\
  Fe I  &   3521.260 &     -0.990 &    0.920 &      47.91& \\
  Fe I  &   3536.560 &      0.120 &    2.880 &       9.38& \\
  Fe I  &   3554.930 &      0.540 &    2.830 &      20.21& \\
  Fe I  &   3558.510 &     -0.630 &    0.990 &      60.21& \\
  Fe I  &   3565.380 &     -0.130 &    0.960 &      73.34& \\
  Fe I  &   3570.100 &      0.150 &    0.920 &      91.51& \\
  Fe I  &   3581.190 &      0.410 &    0.860 &      96.92& \\
  Fe I  &   3585.320 &     -0.800 &    0.960 &      54.52& \\
  Fe I  &   3585.710 &     -1.190 &    0.920 &      42.65& \\
  Fe I  &   3586.110 &      0.170 &    3.240 &       6.64& \\
  Fe I  &   3586.990 &     -0.800 &    0.990 &      52.90& \\
  Fe I  &   3603.200 &     -0.260 &    2.690 &       7.28& \\
  Fe I  &   3608.860 &     -0.100 &    1.010 &      74.12& \\
  Fe I  &   3618.770 &      0.000 &    0.990 &      76.59& \\
  Fe I  &   3621.460 &     -0.020 &    2.730 &      10.91& \\
  Fe I  &   3622.000 &     -0.150 &    2.760 &       7.53& \\
  Fe I  &   3631.460 &     -0.040 &    0.960 &      80.98& \\
  Fe I  &   3640.390 &     -0.110 &    2.730 &       8.22& \\
  Fe I  &   3647.840 &     -0.190 &    0.920 &      77.55& \\
  Fe I  &   3727.619 &     -0.620 &    0.958 &      65.02& \\
  Fe I  &   3743.362 &     -0.780 &    0.990 &      59.53& \\
  Fe I  &   3745.561 &     -0.770 &    0.087 &      97.81& \\
  Fe I  &   3745.900 &     -1.340 &    0.121 &      78.13& \\
  Fe I  &   3758.233 &     -0.020 &    0.958 &      83.49& \\
  Fe I  &   3763.789 &     -0.230 &    0.990 &      75.98& \\
  Fe I  &   3767.192 &     -0.389 &    1.011 &      69.21& \\
  Fe I  &   3786.680 &     -2.220 &    1.010 &       7.82& \\
  Fe I  &   3787.880 &     -0.859 &    1.011 &      55.84& \\
  Fe I  &   3790.093 &     -1.740 &    0.990 &      20.28& \\
  Fe I  &   3812.964 &     -1.030 &    0.958 &      51.49& \\
  Fe I  &   3815.840 &      0.240 &    1.485 &      74.11& \\
  Fe I  &   3820.425 &      0.140 &    0.859 &      97.49& \\
  Fe I  &   3824.444 &     -1.362 &    0.000 &      85.00& \\
  Fe I  &   3825.881 &     -0.030 &    0.915 &      87.53& \\
  Fe I  &   3827.823 &      0.080 &    1.557 &      64.94& \\
  Fe I  &   3840.440 &     -0.510 &    0.990 &      68.58& \\
  Fe I  &   3841.048 &     -0.045 &    1.608 &      57.20& \\
  Fe I  &   3843.260 &     -0.240 &    3.050 &       5.11& \\
  Fe I  &   3849.967 &     -0.870 &    1.011 &      55.15& \\
  Fe I  &   3850.818 &     -1.740 &    0.990 &      19.68& \\
  Fe I  &   3856.372 &     -1.280 &    0.052 &      83.81& \\
  Fe I  &   3859.911 &     -0.710 &    0.000 &     104.34& \\
  Fe I  &   3865.523 &     -0.950 &    1.011 &      52.93& \\
  Fe I  &   3878.018 &     -0.896 &    0.958 &      58.79& \\
  Fe I  &   3878.573 &     -1.360 &    0.087 &      83.55& \\
  Fe I  &   3895.656 &     -1.660 &    0.110 &      71.42& \\
  Fe I  &   3899.707 &     -1.520 &    0.087 &      76.59& \\
  Fe I  &   3902.946 &     -0.440 &    1.558 &      47.77& \\
  Fe I  &   3906.480 &     -2.200 &    0.110 &      44.76& \\
  Fe I  &   3917.180 &     -2.150 &    0.990 &       8.69& \\
  Fe I  &   3920.258 &     -1.740 &    0.121 &      66.89& \\
  Fe I  &   3922.912 &     -1.640 &    0.052 &      74.87& \\
  Fe I  &   3927.920 &     -1.522 &    0.110 &      77.97& \\
  Fe I  &   3930.297 &     -1.491 &    0.087 &      77.30& \\
  Fe I  &   3956.680 &     -0.430 &    2.690 &       7.27& \\
  Fe I  &   3225.790 &      0.380 &    2.400 &      27.36& \\
  Fe I  &   3286.750 &     -0.170 &    2.180 &      21.74& \\
  Fe I  &   3407.460 &     -0.020 &    2.180 &      26.08& \\
  Fe I  &   3440.610 &     -0.670 &    0.000 &      92.40& \\
  Fe I  &   3440.990 &     -0.960 &    0.050 &      82.79& \\
  Fe I  &   3443.880 &     -1.370 &    0.090 &      69.59& \\
  Fe I  &   3475.450 &     -1.050 &    0.090 &      79.44& \\
  Fe I  &   3476.700 &     -1.510 &    0.120 &      63.99& \\
  Fe I  &   3490.570 &     -1.110 &    0.050 &      78.85& \\
  Fe I  &   3497.840 &     -1.550 &    0.110 &      63.13& \\
  Fe I  &   3521.260 &     -0.990 &    0.920 &      47.91& \\
  Fe I  &   3536.560 &      0.120 &    2.880 &       9.38& \\
  Fe I  &   3554.930 &      0.540 &    2.830 &      20.21& \\
  Fe I  &   3558.510 &     -0.630 &    0.990 &      60.21& \\
  Fe I  &   3565.380 &     -0.130 &    0.960 &      73.34& \\
  Fe I  &   3570.100 &      0.150 &    0.920 &      91.51& \\
  Fe I  &   3581.190 &      0.410 &    0.860 &      96.92& \\
  Fe I  &   3585.320 &     -0.800 &    0.960 &      54.52& \\
  Fe I  &   3585.710 &     -1.190 &    0.920 &      42.65& \\
  Fe I  &   3586.110 &      0.170 &    3.240 &       6.64& \\
  Fe I  &   3586.990 &     -0.800 &    0.990 &      52.90& \\
  Fe I  &   3603.200 &     -0.260 &    2.690 &       7.28& \\
  Fe I  &   3608.860 &     -0.100 &    1.010 &      74.12& \\
  Fe I  &   3618.770 &      0.000 &    0.990 &      76.59& \\
  Fe I  &   3621.460 &     -0.020 &    2.730 &      10.91& \\
  Fe I  &   3622.000 &     -0.150 &    2.760 &       7.53& \\
  Fe I  &   3631.460 &     -0.040 &    0.960 &      80.98& \\
  Fe I  &   3640.390 &     -0.110 &    2.730 &       8.22& \\
  Fe I  &   3647.840 &     -0.190 &    0.920 &      77.55& \\
  Fe I  &   3727.619 &     -0.620 &    0.958 &      65.02& \\
  Fe I  &   3743.362 &     -0.780 &    0.990 &      59.53& \\
  Fe I  &   3745.561 &     -0.770 &    0.087 &      97.81& \\
  Fe I  &   3745.900 &     -1.340 &    0.121 &      78.13& \\
  Fe I  &   3758.233 &     -0.020 &    0.958 &      83.49& \\
  Fe I  &   3763.789 &     -0.230 &    0.990 &      75.98& \\
  Fe I  &   3767.192 &     -0.389 &    1.011 &      69.21& \\
  Fe I  &   3786.680 &     -2.220 &    1.010 &       7.82& \\
  Fe I  &   3787.880 &     -0.859 &    1.011 &      55.84& \\
  Fe I  &   3790.093 &     -1.740 &    0.990 &      20.28& \\
  Fe I  &   3812.964 &     -1.030 &    0.958 &      51.49& \\
  Fe I  &   3815.840 &      0.240 &    1.485 &      74.11& \\
  Fe I  &   3820.425 &      0.140 &    0.859 &      97.49& \\
  Fe I  &   3824.444 &     -1.362 &    0.000 &      85.00& \\
  Fe I  &   3825.881 &     -0.030 &    0.915 &      87.53& \\
  Fe I  &   3827.823 &      0.080 &    1.557 &      64.94& \\
  Fe I  &   3840.440 &     -0.510 &    0.990 &      68.58& \\
  Fe I  &   3841.048 &     -0.045 &    1.608 &      57.20& \\
  Fe I  &   3843.260 &     -0.240 &    3.050 &       5.11& \\
  Fe I  &   3849.967 &     -0.870 &    1.011 &      55.15& \\
  Fe I  &   3850.818 &     -1.740 &    0.990 &      19.68& \\
  Fe I  &   3856.372 &     -1.280 &    0.052 &      83.81& \\
  Fe I  &   3859.911 &     -0.710 &    0.000 &     104.34& \\
  Fe I  &   3865.523 &     -0.950 &    1.011 &      52.93& \\
  Fe I  &   3878.018 &     -0.896 &    0.958 &      58.79& \\
  Fe I  &   3878.573 &     -1.360 &    0.087 &      83.55& \\
  Fe I  &   3895.656 &     -1.660 &    0.110 &      71.42& \\
  Fe I  &   3899.707 &     -1.520 &    0.087 &      76.59& \\
  Fe I  &   3902.946 &     -0.440 &    1.558 &      47.77& \\
  Fe I  &   3906.480 &     -2.200 &    0.110 &      44.76& \\
  Fe I  &   3917.180 &     -2.150 &    0.990 &       8.69& \\
  Fe I  &   3920.258 &     -1.740 &    0.121 &      66.89& \\
  Fe I  &   3922.912 &     -1.640 &    0.052 &      74.87& \\
  Fe I  &   3927.920 &     -1.522 &    0.110 &      77.97& \\
  Fe I  &   3930.297 &     -1.491 &    0.087 &      77.30& \\
  Fe I  &   3956.680 &     -0.430 &    2.690 &       7.27& \\
  Fe I  &   4005.242 &     -0.600 &    1.558 &      42.04& \\
  Fe I  &   4045.812 &      0.280 &    1.486 &      78.17& \\
  Fe I  &   4062.441 &     -0.862 &    2.845 &       1.84& \\
  Fe I  &   4063.594 &      0.062 &    1.558 &      67.63& \\
  Fe I  &   4067.978 &     -0.472 &    3.211 &       2.13& \\
  Fe I  &   4071.738 &     -0.008 &    1.608 &      62.82& \\
  Fe I  &   4107.488 &     -0.879 &    2.831 &       1.62& \\
  Fe I  &   4109.802 &     -0.895 &    2.845 &       1.77& \\
  Fe I  &   4118.545 &      0.215 &    3.573 &       4.34& \\
  Fe I  &   4127.608 &     -0.990 &    2.858 &       1.55& \\
  Fe I  &   4132.058 &     -0.675 &    1.608 &      37.12& \\
  Fe I  &   4132.899 &     -1.006 &    2.845 &       1.99& \\
  Fe I  &   4134.678 &     -0.649 &    2.831 &       3.32& \\
  Fe I  &   4143.415 &     -0.204 &    3.047 &       5.36& \\
  Fe I  &   4143.868 &     -0.511 &    1.558 &      45.70& \\
  Fe I  &   4147.669 &     -2.090 &    1.485 &       3.80& \\
  Fe I  &   4153.900 &     -0.321 &    3.396 &       2.30& \\
  Fe I  &   4154.806 &     -0.400 &    3.368 &       2.02& \\
  Fe I  &   4156.799 &     -0.809 &    2.831 &       2.61& \\
  Fe I  &   4175.636 &     -0.827 &    2.845 &       2.42& \\
  Fe I  &   4181.755 &     -0.371 &    2.831 &       6.91& \\
  Fe I  &   4187.039 &     -0.541 &    2.449 &      10.20& \\
  Fe I  &   4187.795 &     -0.530 &    2.425 &      11.44& \\
  Fe I  &   4191.431 &     -0.666 &    2.469 &       6.81& \\
  Fe I  &   4199.095 &      0.156 &    3.047 &      13.61& \\
  Fe I  &   4202.029 &     -0.689 &    1.485 &      43.19& \\
  Fe I  &   4216.184 &     -3.356 &    0.000 &      10.84& \\
  Fe I  &   4222.213 &     -0.914 &    2.449 &       4.35& \\
  Fe I  &   4227.427 &      0.266 &    3.332 &       7.50& \\
  Fe I  &   4233.603 &     -0.579 &    2.482 &       8.14& \\
  Fe I  &   4235.937 &     -0.330 &    2.425 &      17.42& \\
  Fe I  &   4250.120 &     -0.380 &    2.469 &      12.69& \\
  Fe I  &   4250.787 &     -0.713 &    1.558 &      37.95& \\
  Fe I  &   4260.474 &      0.077 &    2.400 &      29.69& \\
  Fe I  &   4271.154 &     -0.337 &    2.449 &      18.47& \\
  Fe I  &   4271.761 &     -0.173 &    1.485 &      65.76& \\
  Fe I  &   4282.403 &     -0.779 &    2.176 &      10.50& \\
  Fe I  &   4325.762 &      0.006 &    1.608 &      65.23& \\
  Fe I  &   4337.046 &     -1.695 &    1.557 &       7.94& \\
  Fe I  &   4375.930 &     -3.020 &    0.000 &      20.68& \\
  Fe I  &   4383.545 &      0.208 &    1.485 &      80.38& \\
  Fe I  &   4404.750 &     -0.147 &    1.557 &      62.82& \\
  Fe I  &   4415.122 &     -0.621 &    1.608 &      43.25& \\
  Fe I  &   4427.310 &     -2.924 &    0.052 &      19.63& \\
  Fe I  &   4430.614 &     -1.728 &    2.223 &       1.47& \\
  Fe I  &   4447.717 &     -1.339 &    2.223 &       3.55& \\
  Fe I  &   4459.118 &     -1.310 &    2.176 &       4.42& \\
  Fe I  &   4461.653 &     -3.200 &    0.087 &      12.01& \\
  Fe I  &   4466.552 &     -0.600 &    2.831 &       3.95& \\
  Fe I  &   4476.019 &     -0.819 &    2.845 &       3.65& \\
  Fe I  &   4489.739 &     -3.930 &    0.121 &       2.18& \\
  Fe I  &   4494.563 &     -1.136 &    2.198 &       5.41& \\
  Fe I  &   4528.614 &     -0.850 &    2.176 &      10.66& \\
  Fe I  &   4531.148 &     -2.130 &    1.485 &       3.77& \\
  Fe I  &   4602.941 &     -2.209 &    1.485 &       3.35& \\
  Fe I  &   4871.318 &     -0.363 &    2.865 &       6.03& \\
  Fe I  &   4872.138 &     -0.567 &    2.882 &       4.27& \\
  Fe I  &   4891.492 &     -0.112 &    2.852 &       9.57& \\
  Fe I  &   4918.994 &     -0.342 &    2.845 &       5.79& \\
  Fe I  &   4920.503 &      0.068 &    2.832 &      14.19& \\
  Fe I  &   4957.299 &     -0.408 &    2.851 &       4.77& \\
  Fe I  &   4957.597 &      0.233 &    2.808 &      22.83& \\
  Fe I  &   4994.130 &     -3.080 &    0.915 &       2.41& \\
  Fe I  &   5006.119 &     -0.638 &    2.832 &       3.06& \\
  Fe I  &   5012.068 &     -2.642 &    0.859 &       7.31& \\
  Fe I  &   5041.756 &     -2.203 &    1.485 &       3.41& \\
  Fe I  &   5049.820 &     -1.420 &    2.279 &       2.90& \\
  Fe I  &   5051.635 &     -2.795 &    0.915 &       4.39& \\
  Fe I  &   5083.339 &     -2.958 &    0.958 &       2.94& \\
  Fe I  &   5123.720 &     -3.068 &    1.011 &       1.46& \\
  Fe I  &   5171.596 &     -1.793 &    1.485 &       9.71& \\
  Fe I  &   5191.455 &     -0.551 &    3.038 &       2.81& \\
  Fe I  &   5192.344 &     -0.421 &    2.998 &       3.66& \\
  Fe I  &   5216.274 &     -2.150 &    1.608 &       3.00& \\
  Fe I  &   5232.940 &     -0.058 &    2.940 &       9.05& \\
  Fe I  &   5266.555 &     -0.386 &    2.998 &       3.87& \\
  Fe I  &   5269.537 &     -1.321 &    0.860 &      59.09& \\
  Fe I  &   5270.356 &     -1.510 &    1.608 &      16.62& \\
  Fe I  &   5324.179 &     -0.240 &    3.211 &       4.85& \\
  Fe I  &   5328.039 &     -1.466 &    0.915 &      47.30& \\
  Fe I  &   5328.532 &     -1.850 &    1.557 &       6.76& \\
  Fe I  &   5429.697 &     -1.879 &    0.958 &      24.62& \\
  Fe I  &   5434.524 &     -2.122 &    1.011 &      14.20& \\
  Fe I  &   5446.917 &     -1.930 &    0.990 &      20.46& \\
  Fe I  &   5455.610 &     -2.091 &    1.011 &      16.05& \\
  Fe I  &   5497.516 &     -2.849 &    1.011 &       2.90& \\
  Fe I  &   5501.465 &     -2.950 &    0.958 &       2.06& \\
  Fe I  &   5506.779 &     -2.797 &    0.990 &       3.73& \\
  Fe I  &   5572.842 &     -0.310 &    3.396 &       1.80& \\
  Fe I  &   5586.756 &     -0.144 &    3.368 &       3.04& \\
  Fe I  &   5615.644 &     -0.140 &    3.332 &       4.57& \\
  Fe I  &   6136.615 &     -1.400 &    2.453 &       2.02& \\
  Fe I  &   6230.723 &     -1.281 &    2.559 &       2.02& \\
  Fe II &   3255.900 &     -2.498 &    0.990 &      38.59& \\
  Fe II &   3277.360 &     -2.191 &    0.990 &      44.91& \\
  Fe II &   3281.300 &     -2.678 &    1.040 &      26.61& \\
  Fe II &   3295.820 &     -2.900 &    1.080 &      15.13& \\
  Fe II &   4233.172 &     -1.809 &    2.583 &       6.20& \\
  Fe II &   4555.893 &     -2.250 &    2.828 &       1.39& \\
  Fe II &   4583.837 &     -1.740 &    2.807 &       5.01& \\
  Fe II &   4923.927 &     -1.206 &    2.891 &      10.61& \\
  Fe II &   5018.430 &     -1.230 &    2.891 &      14.66& \\
  Fe II &   5169.033 &     -1.140 &    2.891 &      20.04& \\
  Fe II &   5316.615 &     -2.020 &    3.153 &       2.05& \\
  Co I  &   3409.170 &     -0.230 &    0.510 &      20.68& \\
  Co I  &   3412.330 &      0.030 &    0.510 &      31.92& \\
  Co I  &   3412.630 &     -0.780 &    0.000 &      31.99& \\
  Co I  &   3443.640 &     -0.010 &    0.510 &      29.51& \\
  Co I  &   3449.440 &     -0.500 &    0.430 &      17.63& \\
  Co I  &   3483.410 &     -1.000 &    0.510 &       4.42& \\
  Co I  &   3502.280 &      0.070 &    0.430 &      50.40& \\
  Co I  &   3502.620 &     -1.240 &    0.170 &      10.61& \\
  Co I  &   3518.350 &      0.070 &    1.050 &       8.81& \\
  Co I  &   3521.570 &     -0.580 &    0.430 &      11.95& \\
  Co I  &   3523.430 &     -0.440 &    0.630 &      16.97& \\
  Co I  &   3526.840 &     -0.620 &    0.000 &      47.06& \\
  Co I  &   3529.020 &     -0.880 &    0.170 &      17.13& \\
  Co I  &   3529.810 &     -0.070 &    0.510 &      31.89& \\
  Co I  &   3842.046 &     -0.770 &    0.923 &       3.79& \\
  Co I  &   3845.461 &     -0.010 &    0.923 &      15.66& \\
  Co I  &   3894.073 &      0.090 &    1.049 &      13.29& \\
  Co I  &   3995.302 &     -0.140 &    0.922 &      12.01& \\
  Co I  &   4118.767 &     -0.470 &    1.049 &       4.71& \\
  Co I  &   4121.311 &     -0.300 &    0.922 &       9.96& \\
  Ni I  &   3232.930 &     -1.010 &    0.000 &      51.74& \\
  Ni I  &   3243.050 &     -1.300 &    0.030 &      40.55& \\
  Ni I  &   3320.250 &     -1.420 &    0.170 &      29.03& \\
  Ni I  &   3369.560 &     -0.660 &    0.000 &      62.52& \\
  Ni I  &   3374.210 &     -1.760 &    0.030 &      23.48& \\
  Ni I  &   3391.040 &     -1.050 &    0.000 &      51.09& \\
  Ni I  &   3392.980 &     -0.540 &    0.030 &      63.76& \\
  Ni I  &   3413.930 &     -1.720 &    0.110 &      20.22& \\
  Ni I  &   3414.760 &     -0.029 &    0.025 &      78.42& \\
  Ni I  &   3423.711 &     -0.730 &    0.212 &      51.30& \\
  Ni I  &   3433.554 &     -0.683 &    0.025 &      61.42& \\
  Ni I  &   3437.270 &     -1.190 &    0.000 &      47.16& \\
  Ni I  &   3452.890 &     -0.910 &    0.109 &      51.28& \\
  Ni I  &   3461.650 &     -0.350 &    0.030 &      76.96& \\
  Ni I  &   3469.480 &     -1.820 &    0.280 &      12.34& \\
  Ni I  &   3472.540 &     -0.810 &    0.110 &      54.17& \\
  Ni I  &   3483.770 &     -1.110 &    0.280 &      37.16& \\
  Ni I  &   3492.950 &     -0.250 &    0.110 &      70.27& \\
  Ni I  &   3500.850 &     -1.280 &    0.170 &      36.24& \\
  Ni I  &   3519.760 &     -1.410 &    0.280 &      24.76& \\
  Ni I  &   3524.530 &      0.010 &    0.030 &      81.65& \\
  Ni I  &   3566.370 &     -0.240 &    0.420 &      56.40& \\
  Ni I  &   3571.860 &     -1.140 &    0.170 &      46.54& \\
  Ni I  &   3597.700 &     -1.100 &    0.210 &      41.83& \\
  Ni I  &   3610.461 &     -1.164 &    0.109 &      45.57& \\
  Ni I  &   3783.524 &     -1.304 &    0.423 &      23.63& \\
  Ni I  &   3807.138 &     -1.220 &    0.423 &      30.00& \\
  Ni I  &   3858.292 &     -0.951 &    0.423 &      42.03& \\
  Ni I  &   4714.408 &      0.260 &    3.380 &       1.11& \\
  Ni I  &   5476.900 &     -0.890 &    1.826 &       3.55& \\
  Cu I  &   3247.530 &     -0.060 &    0.000 &      17.76& \\
  Cu I  &   3273.950 &     -0.360 &    0.000 &       6.48& \\
  Sr II &   4077.709 &      0.158 &    0.000 &      44.69& \\
  Sr II &   4215.519 &     -0.155 &    0.000 &      32.39& \\
   Y II &   3600.740 &      0.280 &    0.180 &       2.45& \\
  Zr II &   3438.230 &      0.420 &    0.090 &       5.38& \\
  Zr II &   3551.960 &     -0.310 &    0.090 &       1.38& \\
  Ba II &   4554.029 &      0.163 &    0.000 &       5.35& \\
  Ba II &   4934.076 &     -0.150 &    0.000 &    $<$3.18& Syn\\
  Be II &   3131.065 &     -0.468 &    0.000 &       $<$0.71& Syn\\
  O  I  &   6300.304 &     -9.819 &    0.000 &       $<$0.40& Syn\\
  O  I  &   7771.950 &      0.358 &    9.147 &       $<$2.00& Syn\\
  K  I  &   7698.974 &     -0.170 &    0.000 &       $<$2.00& Syn\\
  V  I  &   4379.230 &      0.550 &    0.301 &       $<$0.20& Syn\\
  Zn I  &   3302.580 &     -0.057 &    4.030 &       $<$0.40& Syn\\
  Zn I  &   3345.020 &      0.246 &    4.078 &       $<$0.40& Syn\\
  Zn I  &   4722.153 &     -0.390 &    4.030 &       $<$0.20& Syn\\
  Zn I  &   4810.528 &     -0.170 &    4.078 &       $<$0.50& Syn\\
  Eu II &   3819.670 &      0.510 &    0.000 &       $<$0.30& Syn\\
  Eu II &   4129.725 &      0.220 &    0.000 &       $<$0.20& Syn\\
  Eu II &   4205.040 &      0.210 &    0.000 &       $<$0.20& Syn\\
  Pb I  &   4057.790 &     -0.200 &    1.320 &       $<$0.20& Syn\\
\enddata
\tablenotetext{a}{The line analyzed by the spectrum synthesis to determine the upper limit of the abundance is indicated by ``Syn''.}
\end{deluxetable}

\begin{deluxetable}{ccc}
\tablewidth{0pc}
\tablecaption{Heliocentric Radial Velocities for BD+44$^\circ$493 \label{tab:velocity}}
\tablehead{
Epoch & HJD & $v_\mathrm{helio}$ \\
& & ($\mathrm{km\,s^{-1}}$)}
\startdata
2008 Aug 22 & 2454701.14313  & $-150.38$ \\
2008 Oct 4 & 2454743.88522 & $-149.91$  \\
2008 Oct 5 & 2454744.96846 & $-150.41$  \\
2008 Nov 16 & 2454786.74089 & $-149.94$  \\
\enddata
\end{deluxetable}

\begin{deluxetable}{ccccc}
\tablewidth{0pc}
\tablecaption{Interstellar \ion{Na}{1} D lines for {\bd} \label{interstellarD}}
\tablehead{
Component & $\lambda_\mathrm{obs}$ & $\lambda_\mathrm{rest}$ & $v_\mathrm{helio}$ & EW \\
& ({\AA}) & ({\AA}) & ($\mathrm{km\,s^{-1}}$) & (m{\AA}) }
\startdata
\ion{Na}{1} D1 & 5895.58 & 5895.92 & $-1.44$ & 97.8 \\
\ion{Na}{1} D2 & 5889.61 & 5889.92 & $-1.31$ & 133.1
\enddata
\end{deluxetable}

\begin{deluxetable}{cccc}
\tablewidth{0pc}
\tablecaption{Parallax and Proper Motion data for {\bd} \label{hipparcos}}
\tablehead{
& Parallax & Proper Motion & Proper Motion \\
& (mas) & in RA (mas/yr) & in DEC (mas/yr) }
\startdata
results & 4.88 & 117.03 & $-33.12$ \\
error & 1.06 & 1.33 & 0.87
\enddata
\end{deluxetable}

\begin{deluxetable}{cccccc}
\tablewidth{0pc}
\tablecaption{Derived Kinematics for {\bd} \label{tab:kinematics}}
\tablehead{
 $U$ & $V$ & $W$ & $V_{\phi}$ & $e$ & $Z_{\rm max}$ \\
 ({\kms}) & ({\kms}) & ({\kms}) & ({\kms}) & & (kpc) }
\startdata
$-42.63$  &  $-169.13$ &    57.18 &   51.49 & 0.739  & 1.138 
\enddata
\end{deluxetable}

\begin{deluxetable}{cccccccc}
\tablewidth{0pc}
\tablecaption{Photometry data for {\bd} \label{photometry}}
\tablehead{
& $B_T$ & $V_T$ & $J$ & $H$ & $K_s$ & $b\!-\!y$ & $c_1$ }
\startdata
results & 9.745 & 9.167 & 7.659 & 7.269 & 7.202 & 0.451 & 0.239 \\
error & 0.022 & 0.017 & 0.018 & 0.021 & 0.020 & 0.003\tablenotemark{a} & 0.004\tablenotemark{a}
\enddata
\tablenotetext{a}{The errors on $b\!-\!y$ and $c_1$ are random errors
only, and do not include systematic errors.}
\end{deluxetable}

\begin{deluxetable}{ccccc}
\tablewidth{0pc}
\tablecaption{Derived Effective Temperatures \label{Teff}}
\tablehead{
Scale\tablenotemark{a} & System & Color & {\Teff} (giant case) & {\Teff} (dwarf case) }
\startdata
A96,A99,A01 & Johnson & $(B\!-\!V)_0 = 0.493\pm0.028$ & $5749\pm115$ & $5756\pm110$ \\
A96,A99,A01 & TCS & $(V\!-\!K)_0 = 1.793\pm0.026$ & $5426\pm38$ & $5330\pm41$ \\
A96,A99,A01 & TCS & $(J\!-\!H)_0 = 0.374\pm0.028$ & $5166\pm144$ & $5337\pm164$ \\
A96,A99,A01 & TCS & $(J\!-\!K)_0 = 0.416\pm0.027$ & $5381\pm123$ & $5425\pm137$ \\
A96,A99,A01 & Str\"{o}mgren & $(b\!-\!y)_0 = 0.422\pm0.003$\tablenotemark{a} & $5470\pm18$ & $5561\pm19$ \\
GHB09 & Johnson & $(B\!-\!V)_0 = 0.493\pm0.028$ & $5581\pm82$ & $5816\pm132$ \\
GHB09 & 2MASS &$(V\!-\!J)_0 = 1.357\pm0.025$ & $5390\pm44$ & $5280\pm54$ \\
GHB09 & 2MASS &$(V\!-\!H)_0 = 1.735\pm0.027$ & $5434\pm38$ & $5278\pm45$ \\
GHB09 & 2MASS & $(V\!-\!K_s)_0 = 1.794\pm0.026$ & $5423\pm34$ & $5322\pm40$ \\
GHB09 & 2MASS &$(J\!-\!K_s)_0 = 0.437\pm0.027$ & $5505\pm128$ & $5486\pm147$ \\
C10 & Johnson & $(B\!-\!V)_0 = 0.493\pm0.028$ & --- & $5905\pm137$ \\
C10 & 2MASS &$(V\!-\!J)_0 = 1.357\pm0.025$ & --- & $5435\pm50$ \\
C10 & 2MASS &$(V\!-\!H)_0 = 1.735\pm0.027$ & --- & $5386\pm44$ \\
C10 & 2MASS & $(V\!-\!K_s)_0 = 1.794\pm0.026$ & --- & $5430\pm39$ \\
C10 & 2MASS &$(J\!-\!K_s)_0 = 0.437\pm0.027$ & --- & $5621\pm138$ \\
C10 & Str\"{o}mgren & $(b\!-\!y)_0 = 0.422\pm0.003$\tablenotemark{b} & --- & $5643\pm23$
\enddata
\tablenotetext{a}{A96: \citet{Alonso96}; A99: \citet{Alonso99}; A01: \citet{Alonso01}; GHB09:\citet{Gonzalez009}; C10: \citet{Casagrande10}}
\tablenotetext{b}{The photometric error on $b\!-\!y$ includes only
random errors, and does not consider systematic errors.}
\end{deluxetable}

\begin{deluxetable}{cccc}
\tablewidth{0pc}
\tablecaption{Adopted Stellar Parameters for {\bd} \label{parameters}}
\tablehead{
{\Teff} (K) & {\logg} (cgs) & $v_\mathrm{micro}$ (km\,s$^{-1}$) & [Fe/H] }
\startdata
$5430\pm150$ & $3.4\pm0.3$ & $1.3\pm0.3$ & $-3.8$
\enddata
\end{deluxetable}

\begin{deluxetable}{ccrrrrc}
\tablewidth{0pc}
\tablecaption{Abundance Results for {\bd} \label{abundances}}
\tablehead{
Element & Ion & $N_\mathrm{lines}$ & $\log\epsilon(\mathrm{X})$ & [X/H] & [X/Fe] & Note }
\startdata
Li & I & Syn & 1.00 & ... & ... & 6708\,{\AA} \\
Be & II & Syn & $<-1.8$ & ... & ... & 3130\,{\AA} \\
C & CH & Syn & 5.95 & $-2.48$ & +1.35 & \\
N & NH & Syn & 4.40 & $-3.43$ & +0.40 & \\
O & OH & Syn & 6.50 & $-2.19$ & +1.64 & \\
O & I & 1 & $<\!7.20$ & $<\!-1.49$ & $<\!+2.34$ & 6300\,{\AA} \\
O & I & 1 & $<\!7.00$ & $<\!-1.19$ & $<\!+2.14$ & 7772\,{\AA} \\
Na & I & 2 & 2.71 & $-3.53$ & +0.30 & Na D lines \\
Mg & I & 8 & 4.23 & $-3.37$ & +0.46 & \\
Al & I & 1 & 2.06 & $-4.39$ & $-0.56$ & 3962\,{\AA} \\
Si & I & 1 & 4.17 & $-3.34$ & +0.49 & 3906\,{\AA} \\
S & I & 1 & $<\!4.35$ & $<\!-2.77$ & $<\!+1.06$ & 9213\,{\AA} \\
K & I & 1 & $<\!1.95$ & $<\!-3.08$ & $<\!+0.75$ & 7699\,{\AA} \\
Ca & I & 12 & 2.82 & $-3.52$ & +0.31 & \\
Ca & II & 1 & 2.98 & $-3.36$ & +0.47 & 3181\,{\AA} \\
Ca & II & 3 & 3.42 & $-2.92$ & +0.91 & Ca triplet \\
Sc & II & 12 & $-0.39$ & $-3.54$ & +0.29 & \\
Ti & I & 7 & 1.48 & $-3.47$ & +0.36 & \\
Ti & II & 55 & 1.48 & $-3.47$ & +0.36 & \\
V & I & 1 & $<\!0.30$ & $<\!-3.63$ & $<\!+0.20$ & 4379\,{\AA} \\
V & II & 2 & 0.08 & $-3.85$ & $-0.02$ & \\
Cr & I & 6 & 1.44 & $-4.20$ & $-0.37$ & \\
Cr & II & 2 & 1.59 & $-4.05$ & $-0.22$ & \\
Mn & I & 2 & 0.49 & $-4.94$ & $-1.11$ & 4030\,{\AA} \\
Mn & II & 3 & 0.81 & $-4.62$ & $-0.79$ & \\
Fe & I & 158 & 3.67 & $-3.83$ & +0.00 & \\
Fe & II & 11 & 3.68 & $-3.82$ & +0.01 & \\
Co & I & 20 & 1.70 & $-3.29$ & +0.54 & \\
Ni & I & 30 & 2.47 & $-3.75$ & +0.08 & \\
Cu & I & 2 & $-0.55$ & $-4.74$ & $-0.91$ & \\
Zn & I & 1 & $<\!0.90$ & $<\!-3.66$ & $<\!+0.17$ & 3345\,{\AA} \\
Sr & II & 2 & $-1.19$ & $-4.06$ & $-0.23$ & \\
Y & II & 1 & $-1.86$ & $-4.07$ & $-0.24$ & 3601\,{\AA} \\
Zr & II & 2 & $-1.21$ & $-3.79$ & $+0.04$ & \\
Ba & II & 2 & $-2.25$ & $-4.43$ & $-0.60$ & \\
Eu & I & 1 & $<\!-2.90$ & $<\!-3.42$ & $<\!+0.41$ & 3820\,{\AA} \\
Pb & I & 1 & $<\!-0.10$ & $<\!-1.85$ & $<\!+1.98$ & 4058\,{\AA} \\
\enddata
\end{deluxetable}

\begin{deluxetable}{cccrrrcc}
\tabletypesize{\small}
\tablewidth{0pc}
\tablecaption{Abundance Errors for {\bd} \label{errors}}
\tablehead{
Element & Ion & $\sigma_\mathrm{random}$ & $\Delta_{T_\mathrm{eff}}$ & $\Delta_{\log g}$ & $\Delta_{v_\mathrm{micro}}$ & $\sigma_\mathrm{total}$ & $\sigma_\mathrm{total}$  \\
& & & $+150\,\mathrm{K}$ & $+0.3\,\mathrm{dex}$ & $+0.3\,\mathrm{km\,s^{-1}}$ & $\log \epsilon(\mathrm{X})$ & [X/Fe] }
\startdata
Li & I & 0.05 & 0.10 & 0.00 & 0.00 & 0.11 & ... \\
C & CH & 0.10 & 0.30 & $-0.10$ & 0.00 & 0.37 & 0.21 \\
N & NH & 0.25 & 0.30 & $-0.10$ & 0.00 & 0.44 & 0.31 \\
O & OH & 0.10 & 0.35 & $-0.10$ & 0.00 & 0.42 & 0.25 \\
Na & I & 0.09 & 0.14 & $-0.03$ & $-0.03$ & 0.18 & 0.10 \\
Mg & I & 0.05 & 0.15 & $-0.08$ & $-0.04$ & 0.21 & 0.08 \\
Al & I & 0.13 & 0.15 & $-0.01$ & $-0.03$ & 0.20 & 0.14 \\
Si & I & 0.13 & 0.17 & $-0.06$ & $-0.10$ & 0.26 & 0.14 \\
Ca & I & 0.03 & 0.10 & $-0.01$ & $-0.01$ & 0.11 & 0.10 \\
Ca & II & 0.13 & 0.05 & 0.06 & $-0.07$ & 0.18 & 0.15 \\
Ca & II & 0.07 & 0.16 & $-0.08$ & $-0.06$ & 0.23 & 0.27 \\
Sc & II & 0.02 & 0.11 & 0.10 & $-0.02$ & 0.18 & 0.07 \\
Ti & I & 0.05 & 0.16 & 0.00 & 0.00 & 0.17 & 0.08 \\
Ti & II & 0.02 & 0.10 & 0.08 & $-0.06$ & 0.17 & 0.08 \\
V & II & 0.09 & 0.09 & 0.10 & 0.00 & 0.19 & 0.10 \\
Cr & I & 0.05 & 0.16 & $-0.01$ & $-0.01$ & 0.17 & 0.07\\
Cr & II & 0.09 & 0.05 & 0.10 & $-0.02$ & 0.16 & 0.10 \\
Mn & I & 0.09 & 0.18 & 0.00 & 0.00 & 0.20 & 0.11\\
Mn & II & 0.07 & 0.06 & 0.10 & $-0.01$ & 0.16 & 0.08 \\
Fe & I & 0.01 & 0.17 & $-0.03$ & $-0.05$ & 0.19 & ... \\
Fe & II & 0.03 & 0.05 & 0.11 & $-0.02$ & 0.15 & ... \\
Co & I & 0.04 & 0.19 & $-0.01$ & $-0.02$ & 0.20 & 0.06 \\
Ni & I & 0.02 & 0.21 & $-0.04$ & $-0.08$ & 0.25 & 0.06 \\
Cu & I & 0.09 & 0.19 & 0.00 & $-0.01$ & 0.21 & 0.11 \\
Sr & II & 0.09 & 0.11 & 0.09 & $-0.06$ & 0.20 & 0.12 \\
Y & II & 0.13 & 0.11 & 0.10 & 0.00 & 0.22 & 0.15 \\
Zr & II & 0.09 & 0.11 & 0.10 & 0.00 & 0.20 & 0.12 \\
Ba & II & 0.09 & 0.11 & 0.10 & 0.00 & 0.20 & 0.12 \\
\enddata
\end{deluxetable}

\end{document}